\documentclass[12pt]{article}
\usepackage{amsthm, amsmath, natbib}
\usepackage{enumerate}
\usepackage{psfrag,epsf}
\usepackage{graphicx}
\usepackage{multirow}
\usepackage{subcaption}
\usepackage{mathtools}
\usepackage{url} 
\usepackage{titlesec}
\usepackage{hhline}
\titlelabel{\thetitle.\quad}
\usepackage{titlesec}
\usepackage{titletoc}
\usepackage[title,titletoc]{appendix}

\usepackage[doublespacing]{setspace}
\usepackage{tocloft}
\usepackage{color}
\usepackage{caption}

\newcommand{\blind}{1}

\begin{document}
\date{}

\newcommand{\ssection}[1]{\section[#1]{\centering #1}}

\def\spacingset#1{\renewcommand{\baselinestretch}%
{#1}\small\normalsize} \spacingset{1}


\if1\blind
{
  \title{\bf  Robust and Accurate Inference via a Mixture of Gaussian and Student's $t$  Errors}
  \author{Hyungsuk Tak\hspace{.2cm}\\
     University of Notre Dame\\
     \\
     Justin A. Ellis\hspace{.2cm}\\
     Infinia ML\\
     \\
     Sujit K. Ghosh\hspace{.2cm}\\
     North Carolina State University}
  \maketitle
} \fi

\if0\blind
{
  \bigskip
  \bigskip
  \bigskip
  \begin{center}
    {\LARGE\bf An outlier mixture model in Kalman filtering}
\end{center}
  \medskip
} \fi

\bigskip
\begin{abstract}
A Gaussian \textcolor{black}{measurement} error assumption, i.e., an assumption that the data are observed up to Gaussian noise, can bias any parameter estimation in the presence of outliers.  A heavy tailed error assumption based on Student's $t$ distribution  helps reduce the bias. However, it may be less efficient in estimating parameters if the heavy tailed assumption is uniformly applied to all of the data when most of them are normally observed. We propose a mixture error assumption that selectively converts Gaussian errors into  Student's $t$ errors according to latent outlier indicators, leveraging the best of the Gaussian and Student's $t$ errors; a parameter estimation can be not only robust but also accurate. Using simulated hospital profiling data and astronomical  time series of brightness data, we demonstrate the potential for the proposed mixture error assumption to estimate parameters accurately in the presence of outliers.  Supplementary materials are available online.
\end{abstract}

\noindent%
\vfill

\newpage
\spacingset{1} 

\section{Introduction}\label{sec:intro}

An assumption that the data are observed up to Gaussian noise is widely used due to its mathematical and computational simplicity despite its sensitivity to outliers \citep{portnoy2000robust}.  There are two  types of mixture models commonly used to account for outliers. The first type is a mixture of Gaussian distributions. \cite{aitkin1980mixture} propose a mixture of a finite number of  Gaussian distributions with the same mean and different variances\footnote{They also propose a mixture of Gaussian distributions with different means and the same variance or with different means and different variances. However, we focus only on the case with the same mean and different variances as our primary goal is to model errors with mean zero.} so that individual Gaussian errors can have larger variances for outlying observations. \cite{hogg2010data} and \cite{vallisneri2017taming} use this idea to detect and model outliers in analyzing astronomical time series data. This approach, however, fixes the inflation factor of the variance for outliers at a constant (or its estimate) without accounting for its uncertainty.

The second type of model is a scale mixture of Gaussian and inverse-Gamma distributions that converts all of the Gaussian errors into Student's $t$ errors for a robust inference \citep{andrews1974scale, west1984outlier, lange1989robust, peel2000robust, gelman2013bayesian}. This scale mixture has been widely used in various fields such as a robust Kalman-filtering \citep{meinhold1989robustification, giron1994bayesian, roth2013student} and image registration processing \citep{gerogiannis2009mixtures}. However, converting all of the Gaussian errors into Student's $t$ errors does not provide information about outlying observations (i.e., outlier detection) and may result in less efficient parameter estimation when a majority of the errors are concentrated at zero.

We propose a mixture error assumption that selectively converts a Gaussian error into a Student's $t$ error to complement both types of errors. This mixture error can be derived from a mixture of two Gaussian errors with different variances by accounting for the uncertainty of the variance inflation for outliers via a scale mixture of Gaussian and inverse-Gamma distributions. Thus, the proposed mixture error is (marginally) a mixture of two errors that share the same location and scale parameters, while one follows a Gaussian distribution and the other follows a heavy tailed Student's $t$ distribution.  This mixture error takes advantage of Gaussian and Student's $t$ errors, i.e., a mixture error model can be more robust than a Gaussian error model and  lead to more accurate parameter estimation than a Student's $t$ error model.  Also, under the mixture framework it is straightforward to introduce latent outlier indicators that are useful for detecting outliers.

For example, suppose we observe two data sets; one is composed of twenty $i.i.d.$ realizations of N$(0, 1)$ and the other is the same data whose last observation is incorrectly recorded  as $10$. Pretending that the mean of the generative Gaussian distribution is an unknown parameter of interest, we set up a model, $y_i=\mu+\epsilon_i$, where $y_i$ is the $i$-th observation, $\mu$ is the unknown location parameter, and $\epsilon_i$ is an error term. A Gaussian error model sets $\epsilon_i\sim\textrm{N}(0,~ \sigma^2_i)$, where $\sigma_i$ is the known scale of the $i$-th error.  A $t_\nu$ error model assumes $\epsilon_i\sim\sigma_i t_{\nu}$, where $\nu$ denotes the known degrees of freedom.  A mixture error model sets $\epsilon_i\sim\textrm{N}(0,~\sigma_i^2)$ with probability $1-\theta$ and $\epsilon_i\sim\sigma_i t_{\nu}$ otherwise. For simplicity, we set $\sigma_i=1$, $\nu=4$, and $\theta=0.1$ without introducing latent outlier indicators. With an improper flat prior  (Lebesgue) on $\mu$, we fit these three error models on each of the two data  sets. 

\begin{figure}[b!]
\begin{centering}
\includegraphics[scale=0.45]{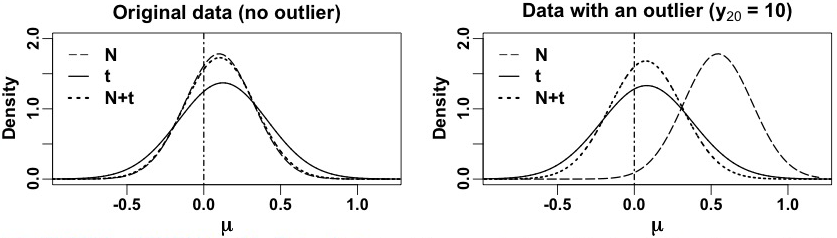}
\caption{The result of fitting three error models, $y_i=\mu+\epsilon_i$, on two data sets, where $\epsilon_i$ follows Gaussian, $t_4$, or their mixture. The original data ($y_i$'s) are 20 realizations of N$(0, 1)$, and the same data with $y_{20}=10$ are used in the right panel. Each curve represents the marginal posterior density of $\mu$ obtained with different errors. The vertical dot-dashed line indicates the generative value, $\mu=0$.  In the first panel, the dotted curve (mixture) intervenes between the other two curves because the mixture error is a weighted average of the other two errors. When there is an outlier, the dotted curve (mixture) puts more mass near $\mu=0$ with less spread than the solid  curve ($t_4$) as shown in the second panel.}
\label{pre_res}
\end{centering}
\end{figure}

In each panel of Figure~\ref{pre_res}, different types of curves denote the marginal posterior densities of $\mu$ obtained with the three different error models; see Appendix~A for details of these marginal posterior densities and their posterior propriety. The generative value, $\mu=0$, is denoted by a vertical dot-dashed line. In the first panel, the dashed  curve (Gaussian) concentrates more on the generative value than the other curves because the data are normally observed without an outlier. The solid  curve ($t_4$) has the widest spread due to the unnecessarily heavy tailed errors  for the normally observed data. Without an  outlier, the dotted  curve (mixture) intervenes between the dashed (Gaussian) and solid ($t_4$) curves, but more closely to the dashed (Gaussian) one. This is because the mixture error is a weighted average of the other two errors and the data are normally observed with no outliers.  In the second panel, the marginal densities of both mixture and $t_4$ error models hardly change in the presence of an outlier \textcolor{black}{with the former (mixture) concentrating more on $\mu=0$. The Gaussian error model, however,}  biases the inference.  This indicates that the parameter estimation with the mixture error can be more accurate than that with the $t_4$ error and more robust than that with the Gaussian error.

\begin{figure}[b!]
\begin{centering}
\includegraphics[scale=0.35]{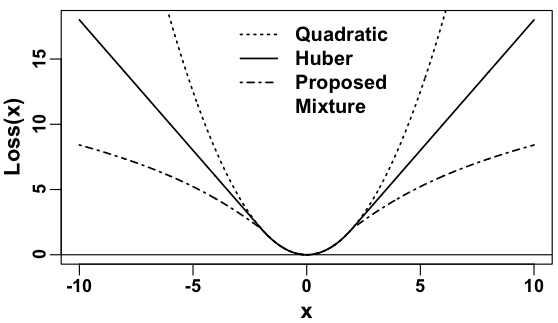}
\caption{\textcolor{black}{Comparison of three loss functions for $k=2$ and $\nu=4$. The quadratic loss function is $\rho(x)=x^2 / 2$. The Huber's loss function is $\rho(x)=x^2 / 2$ if ${\vert x\vert< 2}$, and $\rho(x)=2\vert x\vert - 2$ if $\vert x\vert\ge 2$. The loss function of the proposed mixture error is $\rho(x)=x^2 / 2$ if ${\vert x\vert< 2}$, and $\rho(x)=2.5\log(1 + x^2/4)-2.5\log(2)$+2 if $\vert x\vert\ge 2$. The loss function of the proposed mixture error is more robust to outlying observations than the Huber's one, while both share the quadratic loss for non-outlying observations ($\vert x\vert< 2$).}}
\label{pre_res_loss}
\end{centering}
\end{figure}

\textcolor{black}{Robust statistics has been well documented in the literature, and the proposed mixture error can be represented in Huber's framework \citep{huber1964loss, huber2009robust}. Huber's robust M-estimator is based on a unique loss function defined as $\rho_k(x)=x^2 / 2$ if ${\vert x\vert< k}$, and $\rho_k(x)=k\vert x\vert -k^2 / 2$ if $\vert x\vert\ge k$, where $x$ can be considered as a residual, e.g., $y_i-\mu$ in our simple example.  \cite{huber1964loss} points out that this loss function can be derived from a mixture of Gaussian and Laplace distributions, i.e., $\exp(-\rho_k(x))$, being surprised by the fact that the mixture distribution corresponding to his loss function has much thinner tails than expected. The proposed mixture error may relieve Huber's surprise with heavy tails, resulting in a loss function, $\rho_k(x)=x^2 / 2$ if ${\vert x\vert< k}$, and $\rho_k(x)=\frac{\nu+1}{2}\log(1 + x^2/\nu)-g(k)$ if $\vert x\vert\ge k$, where $g(k)=\frac{\nu+1}{2}\log(1 + k^2/\nu)-k^2/2$. Figure~\ref{pre_res_loss} compares the Huber's loss function with the loss function of the proposed mixture error when $k=2$ and $\nu=4$. It clearly shows that the \textcolor{black}{latter} deals with outlying observations in a more robust way than the former, while both share the quadratic loss for non-outlying observations (${\vert x\vert< 2}$).}

\textcolor{black}{In this article, we propose a Bayesian implementation for the proposed mixture error  for several reasons. First, it is  convenient to make the unknown outlier threshold  ($\vert x\vert\ge k$) random by introducing outlier indicators, e.g.,  $z_i\sim\textrm{Bern}(\theta)$, where $\theta=P(\vert x_i\vert\ge k)$. Second, a Bayesian implementation can be widely applicable because it  enables applying the proposed mixture error  to any existing Gaussian error models in a consistent manner, which we explain later\footnote{Frequentists' robust implementations, on the other hand, may need to develop different estimators for different models, e.g., a homoskedastic ordinary regression model \citep{yohai1987high, gervini2002robust, she2011outlier}, a linear mixed model \citep{copt2006robust}, and an auto-regressive model for regularly-spaced time series data \citep{maronna2006time, bhatia2016robust}.}. Finally, a Bayesian analysis provides a comprehensive set of inferential tools with good properties \citep{berger1985decision}, coherently unifying parameter estimation and uncertainty quantification  via joint posterior distributions.}

In Section~\ref{sec2}, we specify the proposed mixture error in a more general setting and suggest \textcolor{black}{a Bayesian} implementation scheme via a Gibbs sampler, especially for a \textcolor{black}{scenario} where users already have their own Gibbs samplers based on Gaussian error models.  Using simulated \textcolor{black}{(heteroskedastic)} hospital profiling data and \textcolor{black}{(irregularly-spaced)} astronomical  time series of brightness data, we compare the performance of the proposed mixture error with that of  Gaussian, $t_\nu$, and mixture of two Gaussian errors in Section~\ref{sec3}.


\section{A mixture of Gaussian and Student's $t_\nu$ errors}\label{sec2}

A commonly-used $p$-dimensional heteroskedastic Gaussian  error $\boldsymbol{\epsilon}_i$ is defined as
\begin{equation}\label{gaussian_model}
\boldsymbol{\epsilon}_i \sim \textrm{N}_p(\boldsymbol{0},~ \boldsymbol{V}\!_i)
\end{equation}
where $\boldsymbol{0}$ is the vector of zeros with length $p$ and measurement covariance matrix $\boldsymbol{V}\!_i$ is a known or accurately estimated $p\times p$ covariance matrix of datum~$i$ ($i=1, 2, \ldots, n$).  For a more robust error, we may adopt a  heavy tailed $p$-dimensional  $t_\nu$-distribution, i.e.,
\begin{equation}
\boldsymbol{\epsilon}_i\mid \nu \sim\boldsymbol{V}_i^{0.5}t_{p, \nu} \label{t_model}
\end{equation}
where $\boldsymbol{V}_i^{0.5}\boldsymbol{V}_i^{0.5}=\boldsymbol{V}\!_i$ and $t_{p, \nu}$ denotes a $p$-dimensional multivariate $t_\nu$ distribution. Although the degrees of freedom $\nu$ can be treated as either a known constant or an unknown parameter, here we consider $\nu$ as an unknown parameter.  Converting all of Gaussian errors into Student's $t_\nu$ errors improves the robustness to outliers, but can be less efficient in estimating parameters if the heavy tail assumption is redundant for most of the normally observed data. Also, it is challenging to detect outliers in this framework.

Thus, we propose mixing both errors via a latent outlier indicator $z_i$ as follows:
\begin{align}
\begin{aligned}\label{mix_marginal}
\boldsymbol{\epsilon}_i \mid z_i, \nu& \sim \textrm{N}_p\!\left(\boldsymbol{0},~ \boldsymbol{V}\!_i\right)~\textrm{if $z_i=0$,}\\
& \sim\boldsymbol{V}_i^{0.5}t_{p, \nu}~~~~~\textrm{if $z_i=1$,}\\
z_i\mid \theta \sim \textrm{Bernoulli}(\theta),~~ \theta  \sim &\textrm{Beta}(km,~k(1-m)),~~ \nu \sim \textrm{Uniform}(1, 40),
\end{aligned}
\end{align}
where $\theta$ is the probability of being an outlier  (i.e., of using a $t_{p, v}$ error)  and $z_i$ is a  latent outlier indicator that is 1 if datum $i$ is an outlying observation and 0 otherwise. This mixture error in~\eqref{mix_marginal} reduces to the Gaussian error  in~\eqref{gaussian_model} if $\theta=0$ and to the $t_\nu$ error  in~\eqref{t_model} if $\theta=1$ with a Uniform(1, 40) prior on $\nu$.  We put a  Beta($km,~k(1-m)$) prior distribution on $\theta$, whose mean and variance are $m$ and $m(1-m)/(k+1)$, respectively. We  interpret $k$ as the number of pseudo observations that affects the precision of the Beta prior distribution \citep{tak2017data}, and set $m=0.01$  to reflect on our prior belief that the proportion of outlying observations is small.

The resulting posterior inference tends to be sensitive to the shape of the Beta prior on $\theta$, and thus we  conduct  extensive sensitivity analyses for each numerical illustration in Section~\ref{sec3}. These analyses show that the resulting posterior inference becomes similar to that with the $t_\nu$ error if the Beta prior approaches the Uniform(0, 1) prior.  When the data size is large, \textcolor{black}{e.g., $n\ge30$ in the first example of Section~\ref{sec3} and sensitivity analyses,} the resulting inference tends to be more accurate with a large value of $k$ (e.g., $k=n$) because it hinders errors from being heavy tailed unless there is strong evidence for outliers. However, if the data size is small and outlier proportion is large, e.g, $n=20$ with 20\% outliers, the Beta prior with a large value of $k$ may dominate the resulting posterior inference, incorrectly designating Gaussian errors to outliers. In this case, the resulting inference becomes biased as the Gaussian error model does. Therefore, when the data size is small \textcolor{black}{(e.g., $n<30$)} it is desirable to use a Uniform(0, 1) prior on $\theta$ to prevent such biased inference; \textcolor{black}{for the cases of $n=20$ with 20\% and 30\% outliers in our sensitivity analyses, the inference with the proposed mixture error becomes similar to that with the $t_\nu$ error, reducing the bias.}
%


For computational convenience, we re-express $\boldsymbol{V}_i^{0.5}t_{p, \nu}$ in~\eqref{mix_marginal} by a scale mixture of Gaussian and inverse-Gamma distributions, introducing an auxiliary variable $\alpha_i$ as follows:
\begin{align}
\begin{aligned}\label{mixture_model}
\boldsymbol{\epsilon}_i \mid z_i, \alpha_i &\sim \textrm{N}_p(\boldsymbol{0},~ \alpha_i^{z_i}\boldsymbol{V}\!_i),\\
z_i\mid \theta \sim \textrm{Bernoulli}(\theta&),~~ \theta \sim \textrm{Beta}(km,~k(1-m)),\\
\alpha_i \mid \nu\sim \textrm{inverse-Gamma}&(\nu/2,~ \nu/2),~~ \nu \sim \textrm{Uniform}(1, 40).
\end{aligned}
\end{align}
Marginally, \eqref{mixture_model}  is equivalent to~\eqref{mix_marginal}. This mixture error in~\eqref{mixture_model} also reduces to a mixture of two Gaussian errors with the same mean and different variances if $\alpha_i$ is fixed at a constant or at its MLE \citep{aitkin1980mixture, hogg2010data, vallisneri2017taming}; the key difference is whether we account for the uncertainty of $\alpha_i$ or not. 

Any Gaussian error model with~\eqref{gaussian_model} can be converted to the proposed  mixture error model with~\eqref{mixture_model} simply via multiplying $\alpha_i^{z_i}$ by the known variance component $\boldsymbol{V}\!_i$ in~\eqref{gaussian_model}. The  extra cost of using this mixture error is to  account for the uncertainties of the additional unknown parameters, $\boldsymbol{z}$, $\theta$, $\boldsymbol{\alpha}$, and $\nu$ in~\eqref{mixture_model}.  Handling these additional  parameters is not computationally expensive. For example, suppose we have  a Gibbs sampler for a Gaussian error model that adopts~\eqref{gaussian_model}. Multiplying $\alpha_i^{z_i}$ by $\boldsymbol{V}\!_i$ changes the original Gibbs sampler in two ways. First, we replace $\boldsymbol{V}\!_i$ with $\alpha_i^{z_i}\boldsymbol{V}\!_i$ in  the original Gibbs sampler to  update parameters other than $\boldsymbol{z}$, $\theta$, $\boldsymbol{\alpha}$, and $\nu$. This implies that we can keep using the original sampler with a slight modification. Second, we additionally update $\boldsymbol{z}$, $\theta$, $\boldsymbol{\alpha}$,  and $\nu$  at the end of each iteration of the (modified) original Gibbs sampler using their conditional posterior distributions, i.e., for $i=1, 2, \ldots, n$,
\begin{align}
\begin{aligned}\label{additional_step}
z_i\mid \theta, \boldsymbol{\alpha}, \nu, \textrm{other parameters}, \textrm{data}&\sim \textrm{Bernoulli}(p_i),\\
\theta\mid \boldsymbol{z}, \boldsymbol{\alpha}, \nu, \textrm{other parameters}, \textrm{data}&\sim \textrm{Beta}\!\left(km+\sum_{i=1}^n z_i,~k(1-m)+n-\sum_{i=1}^nz_i\right)\!,\\
\alpha_i\mid \boldsymbol{z}, \theta, \nu, \textrm{other parameters}, \textrm{data}&\sim \textrm{inverse-Gamma}\!\left(\frac{\nu+z_i}{2},~ w_i\right)\!,\\
\pi(\nu \mid \boldsymbol{z}, \theta, \boldsymbol{\alpha}, \textrm{other parameters},  \textrm{data}&)  \propto \frac{(\nu / 2)^{n\nu/2}}{\Gamma(\nu/2)^n}\exp\!\left(-\frac{\nu}{2}\sum_{i=1}^n \left(\log(\alpha_i) + \frac{1}{\alpha_i}\right)\right),\\
\end{aligned}
\end{align}
where $\nu\in(1, 40)$. Here, the parameter of the Bernoulli distribution $p_i$  is a proportion of $\theta$-weighted Gaussian densities with the same case-specific mean and different variances, $V_i$ and $\alpha_i V_i$.  The scale parameter of the inverse-Gamma distribution $w_i$ is also case-specific but easy to compute. Since the conditional posterior distribution of $\nu$ is not a standard family distribution, we sample $\nu$ from a Metropolis-Hastings kernel that is invariant to $\pi(\nu \mid \boldsymbol{\alpha}, \theta, \boldsymbol{z}, \textrm{other parameters}, \textrm{data})$. Consequently, these additional updates form a bigger Gibbs loop that encompasses the original Gibbs loop with a slight modification.

Converting a Gaussian error to a mixture error via multiplying $\alpha_i^{z_i}$ by $\boldsymbol{V}\!_i$  extends the original joint posterior distribution incorporating additional parameters $\boldsymbol{z}$, $\boldsymbol{\alpha}$,  $\theta$, and $\nu$. (The extended model does not reduce to the Gaussian error model unless we fix $\theta$ at 0.) Posterior propriety of this extended joint posterior distribution is guaranteed if the original Gaussian error model adopts jointly proper prior distributions for all of the unknown parameters. This is because the additional parameters also have proper prior distributions as specified in~\eqref{mixture_model}. However, it is challenging to prove posterior propriety of the extended joint posterior distribution when the original model adopts jointly improper prior distributions except for some trivial cases such as our toy example in Section~\ref{sec:intro}. This is because marginalizing parameters from the product of the mixtures of Gaussian and $t_{\nu}$ densities is mathematically complicated. In the following numerical illustrations, we use proper prior distributions for unknown parameters to avoid potential posterior impropriety.

\section{Numerical illustrations}\label{sec3}
\textcolor{black}{In our numerical studies, we use R \citep{r2016} to code our model implementations, and all the R codes are available online as a supplementary material.}

\subsection{A two-level Gaussian hierarchical model}\label{ex_hospital}
Here we generate a simulated data set using the data and model of \cite{morris2012shrinkage}, given certain  values of population parameters, and focus on estimating these  parameters in the presence of synthetic outliers. \cite{morris2012shrinkage} analyze  medical profiling data of thirty-one hospitals in New York State using a two-level Gaussian hierarchical model to estimate  random effects regarding the unknown true success rate of coronary artery bypass graft surgery. The original data\footnote{The New York State Department of Health annually releases such data to help people choose hospitals and to improve the quality of medical services (www.health.ny.gov/statistics/diseases/cardiovascular).} are composed of the number of patients in  each hospital who have received the surgery and the number of deaths within a month of the  surgery. \cite{morris2012shrinkage} use an arcsine transformation of the observed success rates to fit their Gaussian hierarchical model; see \cite{tak2017rgbp} and \cite{tak2017data} for analyses via fitting Poisson and Binomial hierarchical models, respectively, without the transformation.  The transformed data are the indices of success rates ($y_i$) that are larger for higher successful surgery rates, and their approximate variances ($V_i$).  The data  are tabulated in Table~\ref{table1}.

To analyze these data, \cite{morris2012shrinkage} set up a two-level Gaussian hierarchical model, i.e., for $i=1, 2, \ldots, 31,$
\begin{equation}
y_i =\mu_i + \epsilon_i~~\textrm{with}~~ \epsilon_i\sim\textrm{N}_1(0,~ V_i)~~\textrm{and}~~\mu_i \mid \beta, A \sim \textrm{N}_1(\beta,~ A),\label{prior_hospital}
\end{equation}
where they assume $V_i$ is known, considering the large number of patients in each hospital, $\mu_i$ denotes the unknown random effect of hospital $i$, and $\beta$ and $A$ are  the unknown mean and variance of the prior (population) distribution for random effects. Our goal is to estimate $\beta$ and $A$ accurately in the presence of outlying observations. Although  \cite{morris2012shrinkage} set an improper joint prior $h(\beta, A)\propto 1$, we adopt a proper one that can mimic their improper choice and guarantee posterior propriety of a mixture error model:
\begin{equation}\label{prior1}
h(\beta, A)\propto \exp\!\left(-\frac{\beta^2}{2\times10^5}\right)  \frac{I_{\{A >0\}}}{(10^5+A)^2},
\end{equation}
where $\beta$ follows a diffuse Gaussian distribution,  $A$ follows a uniform shrinkage prior distribution, $10^5/(10^5+A)\sim\textrm{Uniform}(0, 1)$, and $I_{\{w\}}$ is an indicator function of $w$. This uniform shrinkage prior can approximate the improper flat prior on $A$ with similar frequency coverage properties because $10^5$ is much larger than the $V_i$'s  \citep{tak2016usp}.

\begin{table}[t!]
\begin{center}\caption{The transformed thirty-one hospital profiling data are composed of the indices of success rates ($y_i$), whose values are larger for higher successful surgery rates, and their approximate variances ($V_i$). The values of $y_i$ and $V_i$ are reproduced from Table~4 of \cite{morris2012shrinkage}. We generate simulated data $\boldsymbol{y}^{\textrm{sim}}=\{y_1^{\textrm{sim}}, y_2^{\textrm{sim}}, \ldots, y_{31}^{\textrm{sim}}\}$ via~\eqref{prior_hospital}, i.e., sampling random effects ($\mu_i$'s) given the generative values, $\beta_{\textrm{gen}}=0$ and $A_{\textrm{gen}}=0.722$, and then sampling $\boldsymbol{y}^{\textrm{sim}}$ given the sampled  $\mu_i$'s. For synthetic outliers, we set $y_1^{\textrm{out}}=12.84$ ($= y_1^{\textrm{sim}}+ 4  V_1^{0.5}$), $y_2^{\textrm{out}}=-15.36$ ($=y_2^{\textrm{sim}}-5  V_2^{0.5}$), and $y_{3}^{\textrm{out}}=10.37$ ($=y_3^{\textrm{sim}}+6  V_3^{0.5}$).}\label{table1}
\begin{tabular}{crcrccrcrccrcr}
$i$ &  \multicolumn{1}{c}{$y_i$}  & $V_i$ & \multicolumn{1}{c}{$y_i^{\textrm{sim}}$}  &$~~~$& $i$ & \multicolumn{1}{c}{$y_i$}  & $V_{i}$ & \multicolumn{1}{c}{$y_i^\textrm{sim}$}&$~~~$& $i$ & \multicolumn{1}{c}{$y_i$}  & $V_{i}$ & \multicolumn{1}{c}{$y_i^\textrm{sim}$}\\
\hline
1&  -2.07 & $2.78^2$ & 1.72&& 11&  -1.43 & $1.20^2$ & -0.45 && 21 &-0.08 &$0.96^2$ & 0.02\\
2& -0.22 & $2.76^2$ & -1.56 && 12& 1.56 & $1.14^2$ & -0.55  && 22 & 0.61 &$0.93^2$ & -0.40\\
3& 0.58 & $1.57^2$ &  0.95&& 13&  0.00 & $1.10^2$ & 0.01 && 23 & 2.05 &$0.93^2$ & 1.52\\
4& -1.87 & $1.42^2$ & 0.36 && 14&  0.41 & $1.08^2$ & 2.98 && 24 &  0.57 &$0.91^2$ & -0.49 \\
5& -0.74 & $1.39^2$ & 0.00 && 15&  0.08 & $1.04^2$ & 0.81 && 25 &  1.10 &$0.90^2$ & 0.54\\
6& -1.97 & $1.37^2$ & -1.39 && 16&  -2.15 & $1.03^2$ &  0.24 && 26 &  -2.42 & $0.84^2$ & 0.41\\
7&  -1.90 & $1.36^2$ & 1.64  && 17 & -0.34 & $1.02^2$ & 0.57 &&  27 &  -0.38 & $0.78^2$ & 0.05\\
8&  2.31 & $1.32^2$ & -1.97 && 18& 0.86 &$1.02^2$ & 0.36 &&  28 &  0.07 & $0.75^2$ & -0.01\\
9&  -0.14 & $1.22^2$ &  -1.60 && 19 & 0.01 &$1.01^2$ & 1.34 && 29 &  0.96 & $0.74^2$ & 0.59\\
10&  -1.21 & $1.22^2$ &  -1.09 && 20 &1.11&$0.98^2$ & 1.66 && 30 &  -0.21& $0.66^2$ & -2.03\\
&  & &  && & & & && 31 &  1.14& $0.62^2$ & 0.51\\
\end{tabular}\end{center}
\end{table}

The resulting full posterior density is 
\begin{equation}\label{full_posterior1}
\pi(\boldsymbol{\mu}, \beta, A\mid \boldsymbol{y})\propto h(\beta, A) \prod_{i=1}^{31}\left[f(y_i\mid \mu_i) g(\mu_i\mid \beta, A)\right],
\end{equation}
where $\boldsymbol{\mu}=(\mu_1, \mu_2, \ldots, \mu_{31})$, $\boldsymbol{y}=(y_1, y_2, \ldots, y_{31})$,  the distribution for $h$ is specified in~\eqref{prior1}, and the distributions for $f$ and $g$ are  in~\eqref{prior_hospital}. Posterior propriety holds because we use the proper prior distributions for $\boldsymbol{\mu}$, $\beta$, and $A$.  We sample this full posterior distribution using a Gibbs sampler that iteratively samples the following  conditional posterior distributions:
\begin{equation}\label{hospital_original_gibbs}
\pi_1(\boldsymbol{\mu}\mid \beta, A, \boldsymbol{y}),~\pi_2(\beta\mid \boldsymbol{\mu}, A, \boldsymbol{y}),~\textrm{and}~\pi_3(A\mid \boldsymbol{\mu}, \beta, \boldsymbol{y}).
\end{equation}
We specify  details of these conditional posterior distributions in Appendix~B.1.

\subsubsection{The proposed mixture error model and its implementation}\label{sec311impl}

The Gaussian error in~\eqref{prior_hospital} can be converted to the proposed mixture error simply via multiplying $\alpha^{z_i}_i$ by $V_i$ in~\eqref{prior_hospital} with prior distributions on the additional  parameters, i.e., 
\begin{align}\label{converting}
\begin{aligned}
y_{i}  = \mu_i+\epsilon_i ~~\textrm{with}&~~\epsilon_i\sim \textrm{N}_1(0,~  \alpha_i^{z_i}V_{i}),\\
z_i\mid \theta \sim \textrm{Bernoulli}(\theta),&~~ \theta  \sim \textrm{Beta}(km,~ k(1-m)),\\
\alpha_i\mid \nu\sim \textrm{inverse-Gamma}&(\nu/2,~ \nu/2),~~ \nu \sim \textrm{Uniform}(1, 40),
\end{aligned}
\end{align}
where we set $k=31$ and $m=0.01$; we conduct sensitivity analyses on $k$ and $m$ in Appendix~B.2, including a case where the data are generated with $t_4$ errors. Using this model, we also check the sensitivity according to both data size and outlier proportion in Appendix~B.3. The resulting extended full posterior distribution is 
\begin{align}
\begin{aligned}\label{full_posterior2}
\pi^\ast(\boldsymbol{\mu}, \beta, A, \boldsymbol{z},  \theta, \boldsymbol{\alpha}, \nu\mid \boldsymbol{y}) \propto q(\boldsymbol{z}, \theta, \boldsymbol{\alpha}, \nu) h(\beta, A) \prod_{i=1}^{31}\left[f^\ast(y_i\mid \mu_i, z_i, \alpha_i) g(\mu_i\mid \beta, A)\right],
\end{aligned}
\end{align}
where the distributions for $f^\ast$ and $q$ are  specified in~\eqref{converting}. Posterior propriety holds because prior densities, $q$, $h$, and $g$, are jointly proper. We sample this extended full posterior distribution, using an extended Gibbs sampler that encompasses the original Gibbs sampler. At each iteration, we first sample $\boldsymbol{\mu}$, $\beta$, and $A$ via~\eqref{hospital_original_gibbs} after replacing $V_i$  in $\pi_1(\boldsymbol{\mu}\mid \beta, A, \boldsymbol{y})$  with $\alpha_i^{z_i}V_i$. Then we update the additional parameters using their conditional posterior distributions outlined in~\eqref{additional_step}, i.e., for $i=1, 2, \ldots, 31$,
\begin{align}
\begin{aligned}\label{additional_step2}
z_i\mid \boldsymbol{\alpha}, \theta, \nu, \boldsymbol{\mu}, \beta, A, \boldsymbol{y}&\sim \textrm{Bernoulli}\!\left(\frac{\theta\textrm{N}_1(y_i\mid\mu_i,~ \alpha_i V_i)}{\theta\textrm{N}_1(y_i\mid\mu_i,~ \alpha_i V_i)+(1-\theta)\textrm{N}_1(y_i\mid\mu_i,~ V_i)}\right)\!,\\
\alpha_i\mid \theta, \boldsymbol{z}, \nu, \boldsymbol{\mu}, \beta, A, \boldsymbol{y}&\sim \textrm{inverse-Gamma}\!\left(\frac{\nu+z_i}{2},~ \frac{\nu+z_i\times(y_i-\mu_i)^2/V_i}{2}\right)\!,\\
\end{aligned}
\end{align}
where the notation N$_1(w\mid a, b)$ denotes the Gaussian density of $w$ with mean $a$ and variance $b$, and the conditional  distributions of $\theta$ and $\nu$ are the same as those specified in~\eqref{additional_step}.

We use this extended Gibbs sampler  to obtain the outcomes based on the Gaussian, $t_\nu$, and mixture of two Gaussian errors. Running the extended Gibbs sampler by fixing  $z_i=0$  for all $i$ without updating the additional parameters, $\theta, \boldsymbol{\alpha}$, and $\nu$, results in the outcomes based on the Gaussian error. Similarly, the extended Gibbs sampler that fixes $z_i=1$ for all~$i$ without updating $\theta$ leads to the outcomes based on the $t_\nu$ error. As for the mixture of two Gaussian errors, we assume that $\alpha_j=\alpha$,  following \cite{aitkin1980mixture}, and implement the extended Gibbs sampler after fixing $\alpha$ at its MLE without updating $\nu$; see Appendix~B.1 for details of the MLE.

\subsubsection{Generation and analysis of simulated data}
To compare the performance of the proposed mixture error with that of the Gaussian, $t_\nu$, and mixture of two Gaussian errors, we generate pseudo-data $\boldsymbol{y}^\textrm{sim}\equiv\{y_1^\textrm{sim}, y_2^\textrm{sim}, \ldots, y_{31}^\textrm{sim}\}$ as follows. Using~\eqref{prior_hospital}, we  sample $\boldsymbol{\mu}^{\textrm{sim}}$ given certain generative values, $\beta_{\textrm{gen}}=0$ and $A_{\textrm{gen}}=0.722$, and then  generate $\boldsymbol{y}^{\textrm{sim}}$  given $\boldsymbol{\mu}^{\textrm{sim}}$; we set $\beta_{\textrm{gen}}=0$ as \cite{morris2012shrinkage} assume and set the value of $A_{\textrm{gen}}$ to the the posterior mode\footnote{Using a built-in function, \texttt{density}, of R \citep{r2016}, we set a value that maximizes the estimated density to the posterior mode throughout this article.} of $A$ obtained by fitting the Gaussian error model on $\boldsymbol{y}$.  Table~\ref{table1} exhibits these simulated data. Using $\boldsymbol{y}^{\textrm{sim}}$, we set up two cases: No outlier and three outliers. We consider $\boldsymbol{y}^{\textrm{sim}}$ as the data without outliers in the first case. We make synthetic outliers, replacing $y_1^{\textrm{sim}}$ with $y_1^{\textrm{out}}$ $(=y_1^{\textrm{sim}}$$+4  V_1^{0.5})$, $y_2^{\textrm{sim}}$ with $y_2^{\textrm{out}}$ $(=y_2^{\textrm{sim}}$$-5 V_2^{0.5})$ and $y_3^{\textrm{sim}}$ with $y_3^{\textrm{out}}$ $(=y_3^{\textrm{sim}}$$ + 6  V_3^{0.5})$ for the data in the second case.  We denote this data set with the synthetic outliers by $\boldsymbol{y}^{\textrm{out}}\equiv\{y_1^{\textrm{out}}, y_2^{\textrm{out}}, y_3^{\textrm{out}}, y_4^{\textrm{sim}}, \ldots, y_{31}^{\textrm{sim}}\}$.

We fit the four error models on each of the two data sets, $\boldsymbol{y}^{\textrm{sim}}$ and $\boldsymbol{y}^{\textrm{out}}$. For each error model, we implement the extended Gibbs sampler  by independently running thirty Markov chains  each for 1,050,000 iterations,  discarding the first 50,000 as burn-in iterations.  We thin each Markov chain by a factor of ten, i.e., from length 1,000,000 to 100,000, and we combine these thirty (thinned) Markov chains to summarize the sampling results; see  Appendix~B.4 for details of  Markov chain convergence diagnostics.

\begin{figure}[b!]
\begin{centering}
\includegraphics[scale=0.45]{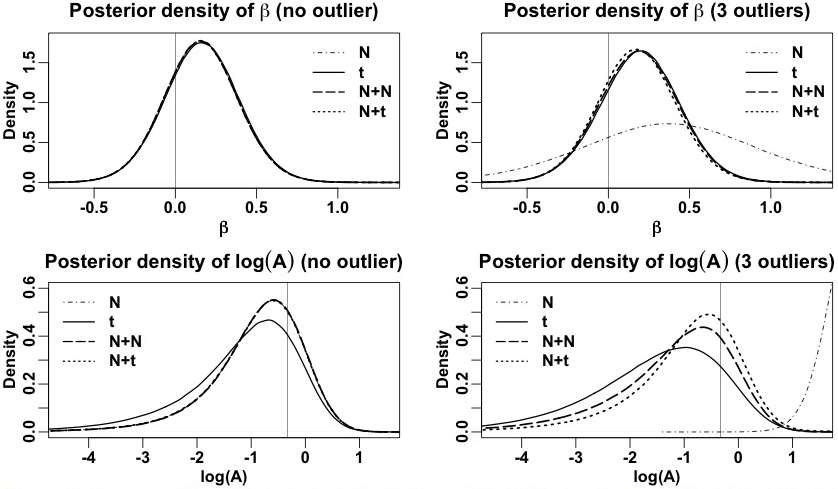}
\caption{Each panel shows posterior densities of $\beta$ (top panels) or those of $\log(A)$ (bottom panels) obtained by fitting four error models on $\boldsymbol{y}^{\textrm{sim}}$ (left) and on $\boldsymbol{y}^{\textrm{out}}$ (right). The generative values, $\beta_{\textrm{gen}}$ and $\log(A_{\textrm{gen}})$, are denoted by vertical lines. Without outliers, all of the density curves for $\beta$ in the top-left panel are indistinguishable, while  the solid  density  curve  ($t$) for $\log(A)$ in the bottom-left panel has the widest spread due to the redundant heavy tailed error assumption. With the  outliers, the dotted curve (proposed mixture) for $\log(A)$ in the bottom-right panel  puts more mass near $\log(A_{\textrm{gen}})$ than the others.}
\label{figure_hospital}
\end{centering}
\end{figure}

Figure~\ref{figure_hospital} displays the sampling results; the upper panels  display the marginal posterior densities of $\beta$ obtained by fitting the four  error models on $\boldsymbol{y}^{\textrm{sim}}$ (left panel) and on $\boldsymbol{y}^{\textrm{out}}$ (right panel), and the bottom panels exhibit those of $\log(A)$. The vertical lines represent the generative values, $\beta_{\textrm{gen}}$ and $\log(A_{\textrm{gen}})$. Without outliers, the four curves for $\beta$ in the top-left panel are  indistinguishable, but the solid  curve~($t_\nu$) for $\log(A)$ in the bottom-left panel has a wider spread than the others. This is because there is no outlying observation and thus the heavy tailed error assumption is unnecessary. With the synthetic outliers, the shape and location of the dot-dashed curves (Gaussian) for both parameters change drastically   as shown in the top- and bottom-right panels. This shows Gaussian error's  sensitivity to outliers. On the other hand, the shape and location of the solid ($t_\nu$), dashed (Gaussian mixture), or dotted (proposed mixture) curve for~$\beta$ hardly change even with the outliers. Comparing these three robust errors in the bottom-right panel, we notice that the dotted curve (proposed mixture)   concentrates more on $\log(A_{\textrm{gen}})$ than the others.

\begin{figure}[b!]
\begin{centering}
\includegraphics[scale=0.45]{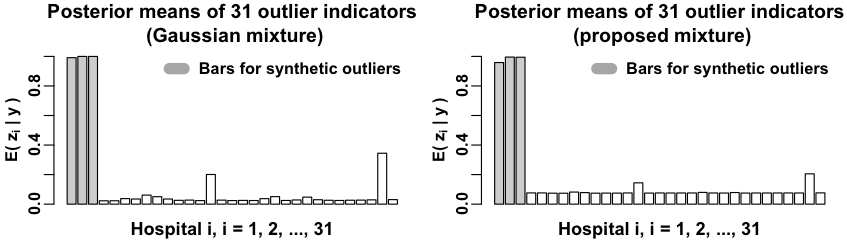}
\caption{Each panel shows posterior means of outlier indicators ($z_i$'s) obtained by fitting the Gaussian mixture error model (left panel) and proposed mixture error model (right panel) on $\boldsymbol{y}^{\textrm{out}}$. The height of each bar represents the average of three million posterior samples of $z_i$ and the horizontal axis indicates hospital $i$ ($i=1, 2, \ldots, 31$). Both models correctly detect the synthetic outliers, $y_{1}^{\textrm{out}}$, $y_{2}^{\textrm{out}}$, and $y_{3}^{\textrm{out}}$. The Gaussian mixture error model works better in designating non-inflated Gaussian errors to normally observed data, although it assigns inflated Gaussian errors to $y_{14}^{\textrm{sim}}$ and $y_{30}^{\textrm{sim}}$ more often.}
\label{figure_hospital_indicator}
\end{centering}
\end{figure}

The mixture framework can provide a functionality to detect outliers via latent outlier indicators, i.e., $z_i$'s. Figure~\ref{figure_hospital_indicator} displays the posterior means of these outlier indicators obtained by fitting the Gaussian mixture error model (left) and proposed mixture error model (right) on $\boldsymbol{y}^{\textrm{out}}$; the height of each bar indicates the average of three million posterior samples of $z_i$. Both models produce posterior means of the first three outlier indicators much higher than the others, correctly detecting the synthetic outliers; in practice it may be desirable  to investigate why these  are considered as outliers. Also, most bars in the first panel have near zero heights while those in the second panel do not, which implies that the Gaussian mixture error model works  better in designating Gaussian errors to normally observed data. However, it designates inflated Gaussian errors to both $y_{14}^{\textrm{sim}}=2.98$ and $y_{30}^{\textrm{sim}}=-2.03$ more often than the proposed mixture error model does. Since relatively large values of $y_{14}^{\textrm{sim}}$ and $y_{30}^{\textrm{sim}}$ are down-weighted more often in the Gaussian mixture error model,  its estimate of the unknown variance component $A$ is likely to be smaller than the one obtained by the proposed mixture error model.  Thus, as shown in the bottom-right panel of Figure~\ref{figure_hospital}, the distribution of $\log(A)$ from the proposed mixture error model puts more mass at larger values of $\log(A)$ than that from the Gaussian mixture error model.


\begin{table}[b!]
\begin{center}\caption{Numerical summaries of the sampling results obtained by fitting the four error models on $\boldsymbol{y}^{\textrm{out}}$. First, we compute the average of 100,000 posterior samples of $\beta$ or $\log(A)$ for each of thirty  Markov chains. The listed posterior mean is the mean of these thirty averages and the Monte Carlo error in the parentheses is the standard deviation of these thirty averages. The bias is the absolute difference between the posterior mean and the generative value. The (Monte Carlo estimate of the) MSE is the bias squared plus the Monte Carlo error squared,  and the MSE ratio is the MSE obtained with the Gaussian, $t_\nu$, or Gaussian mixture error model divided by that obtained with the proposed mixture error model. The 95\% posterior interval (P.I.) is based on 0.025 and 0.975 quantiles of  the combined three million posterior samples. The CPU time in seconds is averaged over the CPU times for the thirty runs. The proposed mixture error model outperforms the other error models in terms of bias, MSE, and 95\% P.I. though it takes more CPU time\textcolor{black}{, as the numbers in bold font indicate}.}\label{hospital_res_table}
\begin{tabular}{ccrcrccr}
&   & Posterior mean & & \multicolumn{1}{c}{MSE} & & Length & \multicolumn{1}{c}{CPU}\\
& Error  &\multicolumn{1}{c}{\small (Monte Carlo error)} &  Bias  & \multicolumn{1}{c}{ratio} & 95\% P.I. & of P.I. & \multicolumn{1}{c}{time}\\
\hline
& N  & 0.376 (0.00068) & 0.376  & \textbf{4.56} & (-0.767, 1.519) & 2.286  & 28\\
\multirow{2}{*}{$\beta$}&$t_\nu$& 0.194 (0.00116) & 0.194  & \textbf{1.22} & (-0.303, 0.682) & 0.985 & 53\\
&N+N& 0.186 (0.00101) & 0.186   & \textbf{1.12} & (-0.302, 0.668) & 0.970 & 48\\
&N$+t_\nu$&  0.176 (0.00086) & \textbf{0.176}   & -  & (-0.305, 0.662) & \textbf{0.967} & \textbf{74}\\\\
& N   & 2.078 (0.00163) &  2.404   & \textbf{13.93} & (1.140, 2.941) & 1.801 & 28\\
\multirow{2}{*}{$\log(A)$}&$t_\nu$ & -1.589 (0.01673) & 1.263   & \textbf{3.85} & (-4.745, 0.374) & 5.119 & 53\\
&N+N & -1.232 (0.01653) & 0.907  & \textbf{1.98} & (-4.201, 0.437)  &  4.638 & 48\\
&N$+t_\nu$ & -0.969 (0.01367) & \textbf{0.644}  & - & (-3.663, 0.592)  &  \textbf{4.255} &  \textbf{74}\\
\end{tabular}\end{center}
\end{table}

To compare the estimation accuracy numerically in the presence of outliers, we summarize the sampling results of $\beta$ and $\log(A)$ in Table~\ref{hospital_res_table} that are obtained by fitting the four error models on $\boldsymbol{y}^{\textrm{out}}$. We list the posterior mean, its Monte Carlo  error and bias, mean-squared error (MSE) ratio, 95\% posterior interval and its length, and the CPU time in seconds; see the caption of Table~\ref{hospital_res_table} for details of their definitions.   With the synthetic outliers,  the proposed mixture error model results in smaller bias, smaller MSE, and shorter 95\% posterior interval for both parameters than the other error models \textcolor{black}{as highlighted in bold font. However, it} takes 1.54 times more CPU time than  the Gaussian mixture error model because it accounts for the uncertainty of variance inflation, i.e., $\boldsymbol{\alpha}$.

\subsection{A state-space model of an Ornstein-Uhlenbeck process}\label{ex_macho}

\begin{figure}[b!]
\begin{centering}
\includegraphics[scale=0.45]{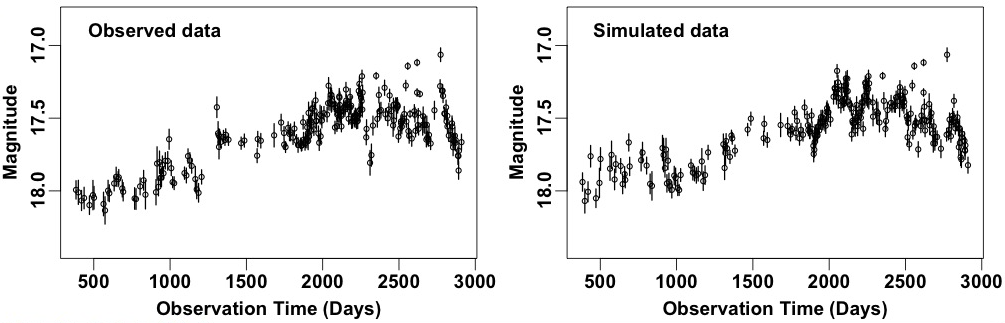}
\caption{The R-band time series data of MACHO source 70.11469.82 in the left panel are composed of 242 magnitudes (astronomical logarithmic measure of brightness) denoted by empty circles and their measurement standard deviations denoted by the half lengths of vertical lines around the empty circles. We generate a simulated data set shown in the right panel by mimicking the observed data as meticulously as possible; see Section~\ref{macho_sim} for details.}
\label{macho_data}
\end{centering}
\end{figure}

We analyze irregularly observed time series data of the brightness of a MACHO (Massive Compact Halo Objects) quasar\footnote{http://www.astro.yale.edu/mgeha/MACHO/70.11469.82.html}  that is a highly luminous galaxy with an actively accreting supermassive  black hole at the center \citep{geha2003variability}. The brightness time series data of MACHO source 70.11469.82 are irregularly observed via an R-band optical filter on 242 nights for 7.5 years since 1992. The data are composed of the magnitudes, an astronomical logarithmic measure of brightness, and their reported measurement standard deviations. The left panel of Figure~\ref{macho_data}  denotes the magnitudes  by empty circles and their measurement standard deviations  by the half lengths of vertical lines around the empty circles. 

 
We use the notation $\boldsymbol{t}=\{t_1, t_2, \ldots, t_{n}\}$ to denote the observation times and  $\boldsymbol{y}=\{y_1, y_2, \ldots, y_{n}\}$ to denote the observed magnitudes ($n=242$).  In analyzing the photometric data,  the  reported measurement variances denoted by $\boldsymbol{V}=\{V_1, V_2, \ldots, V_{n}\}$ are typically assumed to be known \citep{kelly2007some}. We also assume that the latent magnitudes denoted by $\boldsymbol{Y}(\boldsymbol{t})=\{Y(t_1), Y(t_2), \ldots, Y(t_{n})\}$ have generated the observed data $\boldsymbol{y}$ with heteroskedastic Gaussian errors, i.e., for $i=1, 2, \ldots, 242,$
\begin{equation}\label{lik}
y_{i}= Y(t_i)+\epsilon_i~~\textrm{with}~~\epsilon_i\sim \textrm{N}_1(0,~  V_{i}).
\end{equation}

We assume that the latent magnitudes $\boldsymbol{Y}(\boldsymbol{t})$ are the values on a latent continuous-time curve  that is a realization of an Ornstein-Uhlenbeck (O-U) process \citep{kelly2009variations}, i.e., a Gaussian process with Mat$\acute{\textrm{e}}$rn$(1/2)$ kernel \citep{williams2006gaussian}. Many astrophysicists have empirically demonstrated that the ${\textrm{O-U}}$ process describes  stochastic variability of quasar light curves well \citep{kelly2009variations, kozlowski2010quantifying, macleod2010modeling}. The O-U process is defined by the following stochastic differential equation:
\begin{equation}\label{stochastic}
d Y(t) =  - \frac{1}{\tau}\big(Y(t) - \mu\big)d t +\sigma d B(t),
\end{equation}
where $\mu$ and $\sigma$ are  the overall mean and short-term variability of the process on the magnitude scale, respectively, $\tau$ is a timescale in days,  and $B(t)$ is standard Brownian motion. Our goal is to estimate $\sigma$ and $\tau$ accurately because these are known to be associated with physical properties of quasars; for example, luminosity and mass of a quasar's central black hole are negatively correlated with $\sigma$ but positively correlated with $\tau$  \citep{kelly2009variations, macleod2010modeling}. Thus both $\sigma$ and $\tau$ can be used to classify quasars \citep{kozlowski2010quantifying}. The solution of~\eqref{stochastic} provides Gaussian prior distributions of the latent magnitudes:
\begin{align} \label{OU}
\begin{aligned}
Y(t_1)\mid \mu, \sigma^2, \tau &~\sim~ \textrm{N}_1\!\left(\mu,~ \frac{\tau\sigma^2}{2}\right)\!, ~\textrm{and for}~ i=2, 3, \ldots, 242,\\
~~Y(t_i)\mid Y(t_{i-1}), \mu, \sigma^2, \tau &~\sim~ \textrm{N}_1\!\left(\mu + a_i\big(Y(t_{i-1}) - \mu\big) ,~\frac{\tau \sigma^2}{2}(1-a^2_i)\right)\!,
\end{aligned}
\end{align}
where $a_i\equiv \exp(-(t_i - t_{i-1})/\tau)$ is a shrinkage factor that depends on the observation  cadence and $\tau$. Following \cite{tak2016timedelay}, we adopt independent, weakly informative, and proper prior distributions for the O-U parameters, $\mu, \sigma^2$, and $\tau$, i.e., 
\begin{align}\label{hyp}
\begin{aligned}
\mu\sim \textrm{Uniform}(-30, 30),~ \sigma^2 \sim \textrm{inverse-Gamma}(1, 10^{-7}),~ \tau \sim \textrm{inverse-Gamma}(1, 1).
\end{aligned}
\end{align}

The resulting full posterior density of the unknown parameters is proportional to the product of   probability densities of the data and parameters, i.e.,
\begin{align}\label{fullpost}
\begin{aligned}
\pi(\boldsymbol{Y}(\boldsymbol{t}), \mu, \sigma^2, \tau\mid \boldsymbol{y}) \propto &~ h(\mu, \sigma^2, \tau) \prod_{i=1}^{242} f(y_i\mid Y(t_i), V_i)\\
&\times g(Y(t_1)\mid \mu, \sigma^2, \tau)\prod_{i=2}^{242} g(Y(t_i)\mid Y(t_{i-1}), \mu, \sigma^2, \tau),
\end{aligned}
\end{align}
where the distributions of $f$, $g$, and $h$ are specified in~\eqref{lik}, \eqref{OU}, and \eqref{hyp}, respectively. The full posterior distribution is proper because all of the prior distributions are jointly proper. We sample this full posterior distribution using a Gibbs sampler specified in Appendix~C.1 that iteratively samples the following four conditional posterior distributions:
\begin{align}
\begin{aligned}\label{macho_original_gibbs}
\pi_1(\boldsymbol{Y}(\boldsymbol{t}) \mid \mu, \sigma^2, \tau, \boldsymbol{y}),~~\pi_2( \mu \mid \boldsymbol{Y}(\boldsymbol{t}),  \sigma^2, \tau, \boldsymbol{y}),\\
\pi_3( \sigma^2 \mid \boldsymbol{Y}(\boldsymbol{t}),  \mu, \tau, \boldsymbol{y}),~~\pi_4(\tau \mid \boldsymbol{Y}(\boldsymbol{t}),  \mu, \sigma^2, \boldsymbol{y}).
\end{aligned}
\end{align}

\subsubsection{The proposed mixture error model and its implementation}

To convert  Gaussian errors into  mixture errors, we multiply $\alpha_i^{z_i}$ by $V_i$ in~\eqref{lik} with independent prior distributions on the additional parameters, i.e., for $i=1, 2, \ldots, 242$,
\begin{align}\label{converting2}
\begin{aligned}
&~~y_{i}= Y(t_i)+\epsilon_i ~~\textrm{with}~~ \epsilon_i\sim \textrm{N}_1(0,~  \alpha^{z_i}_iV_{i}),\\
z_i&\mid \theta \sim \textrm{Bernoulli}(\theta),~~ \theta \sim \textrm{Beta}(km,~ k(1-m)),\\
\alpha_i \mid \nu &\sim \textrm{inverse-Gamma}(\nu/2,~ \nu/2),~~ \nu \sim \textrm{Uniform}(1, 40),
\end{aligned}
\end{align}
where $k=242$ and $m=0.01$; see Appendix~C.2 for sensitivity analyses on $k$ and $m$, including a case where we generate another data set with $t_4$ errors. The full posterior distribution in~\eqref{fullpost} is extended to
\begin{align}\label{fullpost2}
\begin{aligned}
\pi^\ast(\boldsymbol{Y}(\boldsymbol{t}), \mu, \sigma^2, \tau, \boldsymbol{z}, \theta, \boldsymbol{\alpha}, \nu\mid \boldsymbol{x}) \propto &~ q(\boldsymbol{z}, \theta, \boldsymbol{\alpha}, \nu) h(\mu, \sigma^2, \tau) \prod_{i=1}^{242} f^\ast(y_i\mid Y(t_i), z_i, \alpha_i)\\
&\times g(Y(t_1)\mid \mu, \sigma^2, \tau)\prod_{i=2}^{242} g(Y(t_i)\mid Y(t_{i-1}), \mu, \sigma^2, \tau),
\end{aligned}
\end{align}
where the distributions of $q$ and $f^\ast$ are defined in~\eqref{converting2}. The extended full posterior distribution is also proper because the prior densities, $q$, $h$, and $g$, are jointly proper. An extended  Gibbs sampler to sample~\eqref{fullpost2} keeps using the original Gibbs sampler, iteratively sampling  $\boldsymbol{Y}(\boldsymbol{t})$, $\mu$, $\sigma^2$, and  $\tau$ using~\eqref{macho_original_gibbs} after replacing $V_i$ in $\pi_1(\boldsymbol{Y}(\boldsymbol{t}) \mid \mu, \sigma^2, \tau)$ with $\alpha_i^{z_i}V_i$. At the end of each iteration of the modified original Gibbs sampler,  we update $\boldsymbol{z}$, $\theta$,  $\boldsymbol{\alpha}$, and $\nu$ using their conditional posterior distributions, i.e., for $i=1, 2, \ldots, n$, 
\begin{align}\label{additional_update}
\begin{aligned}
z_i \mid \theta, \boldsymbol{\alpha}, \nu &\sim \textrm{Bernoulli}\!\left(\frac{\theta \times\textrm{N}_1(y_i \mid Y(t_i), \alpha V_i)}{\theta \times \textrm{N}_1(y_i \mid Y(t_i), \alpha V_i)+(1-\theta)\times \textrm{N}_1(y_i \mid Y(t_i), V_i) }\right)\!, \\
 \alpha_i\mid \theta, \boldsymbol{z}, \nu  &\sim \textrm{inverse-Gamma}\!\left(\frac{\nu+z_i}{2},~\frac{\nu + z_i\times(y_i - Y(t_i))^2 /V_i}{2}\right),
\end{aligned}
\end{align}
and the conditional posterior distributions of $\theta$ and $\nu$ are specified in~\eqref{additional_step}. We suppress conditioning on $\boldsymbol{Y}(\boldsymbol{t}), \mu, \sigma^2, \tau,$ and $\boldsymbol{y}$ in~\eqref{additional_update}.

We use this extended Gibbs sampler to obtain the outcomes based on Gaussian, $t_\nu$, and mixture of two Gaussian errors. For the Gaussian error model, we set $z_i=0$ for all~$i$ without updating $\theta$, $\boldsymbol{\alpha}$, and $\nu$. Similarly,  for the $t_\nu$ error model, we fix $z_i=1$ for all~$i$ and do not update $\theta$. Following  \cite{vallisneri2017taming}, we fix $\alpha_i$ at an arbitrarily large constant, $10^2$, for the Gaussian mixture error model. 

\subsubsection{Generation and analysis of simulated data of MACHO  70.11469.82}\label{macho_sim}

To check the effect of outliers on estimating the O-U parameters, we generate a simulated data set, mimicking the original data of MACHO  70.11469.82 as meticulously as possible. First,  we fit the proposed mixture error model on the original data $\boldsymbol{y}$ and remove seven data points whose posterior means of outlier indicators are greater than 0.3, considering that most of the posterior means are about 0.02. These removed values are $y_{155}$, $y_{163}$,  $y_{189}$, $y_{191}$, $y_{199}$, $y_{200}$, and $y_{217}$. Next, we fit a Gaussian error model on the data without the seven  observations and compute the posterior modes of $\mu$, $\sigma^2$, and $\tau$ that are 17.667, 0.018$^2$, and 284.066, respectively, based on one-half million posterior samples. Treating these as generative values, i.e., $\mu_{\textrm{gen}}=17.667$, $\sigma^2_{\textrm{gen}}=0.018^2$, and $\tau_{\textrm{gen}}=284.066$, we start simulating data, i.e., we generate $\boldsymbol{Y}^{\textrm{sim}}(\boldsymbol{t})$ from~\eqref{OU} and  then generate $\boldsymbol{y}^{\textrm{sim}}=\{y_1^{\textrm{sim}}, \ldots, y_{242}^{\textrm{sim}}\}$ from~\eqref{lik} given the sampled $\boldsymbol{Y}^{\textrm{sim}}(\boldsymbol{t})$. Finally, we recover the seven outliers by setting $y^{\textrm{sim}}_{155}=y_{155}$, $y^{\textrm{sim}}_{163}=y_{163}$, $y^{\textrm{sim}}_{189}=y_{189}$, $y^{\textrm{sim}}_{191}=y_{191}$, $y^{\textrm{sim}}_{199}=y_{199}$, $y^{\textrm{sim}}_{200}=y_{200}$, and $y^{\textrm{sim}}_{217}=y_{217}$. This process produces one simulated data set and we repeat this process a million times and choose one that gives the smallest sum of weighted absolute differences defined as $\sum_{i=1}^{242}\vert y_i - y_i^{\textrm{sim}}\vert/V_i^{0.5}$. The simulated data are plotted in the second panel of Figure~\ref{macho_data}.

We fit the four error models on both $\boldsymbol{y}^{\textrm{sim}}$ and $\boldsymbol{y}$. For each error model, we independently run thirty Markov chains each with length 550,000 and discard the first 50,000  as burn-in iterations. We thin each Markov chain from length 500,000 to 100,000. We display and summarize the sampling results using the combined three million posterior samples of each parameter for both simulated and real data analyses; see  Appendix~C.3 for details of Markov chain convergence diagnostics.




\begin{figure}[b!]
\begin{centering}
\includegraphics[scale=0.43]{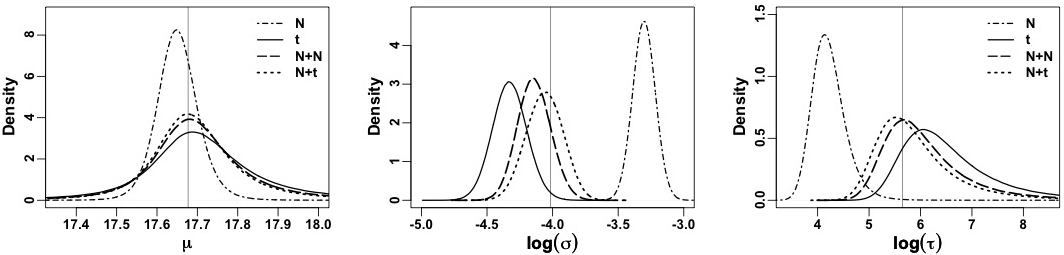}
\caption{The posterior distributions of $\mu$, $\log(\sigma)$, and $\log(\tau)$ (from the left panel) obtained with Gaussian (dot-dashed), $t_\nu$ (solid), Gaussian mixture (dashed), and proposed mixture (dotted) error models. The generative values, $\mu_{\textrm{gen}}$, $\log(\sigma_{\textrm{gen}})$ and $\log(\tau_{\textrm{gen}})$, are denoted by vertical lines. The proposed mixture error model results in  posterior distributions of the parameters of interest, $\log(\sigma)$ and $\log(\tau)$, that put more mass near the generative values than the other error models.}
\label{figure_macho_sim}
\end{centering}
\end{figure}

Figure~\ref{figure_macho_sim} exhibits the posterior distributions of $\mu$, $\log(\sigma)$, and $\log(\tau)$ obtained by fitting the Gaussian (dot-dashed curve), $t_\nu$ (solid curve), Gaussian mixture (dashed curve), and proposed mixture (dotted curve) error models on $\boldsymbol{y}^{\textrm{sim}}$. The vertical lines indicate the generative values, $\mu_{\textrm{gen}}$, $\log(\sigma_{\textrm{gen}})$, and $\log(\tau_{\textrm{gen}})$. In estimating the location parameter $\mu$ in the first panel, the $t_\nu$, Gaussian and proposed mixture error models produce  posterior distributions of $\mu$ that have a wider spread but concentrate closer to $\mu_{\textrm{gen}}$ than the Gaussian error model. In the second panel, the mode of the posterior distribution  of $\log(\sigma)$ obtained with  Gaussian error   is much larger than $\log(\sigma_{\textrm{gen}})$ because the short-term variability $\sigma$ is anticipated to vastly increase to account for the outliers under the Gaussian error assumption. In the third panel, the opposite occurs for the posterior distribution of $\log(\tau)$ obtained with Gaussian error because of the negative association between $\sigma$ and $\tau$ a posteriori \citep{kelly2009variations, macleod2010modeling}. Thus, the Gaussian error assumption leads to severe biases for the parameters of interest, $\sigma$ and $\tau$, in the presence of outliers. When it comes to the comparison between the robust choices, the posterior distributions of the three parameters obtained by the proposed mixture error model puts more mass near the generative values than those obtained by the $t_\nu$ and Gaussian mixture error models.

\begin{figure}[b!]
\begin{centering}
\includegraphics[scale=0.45]{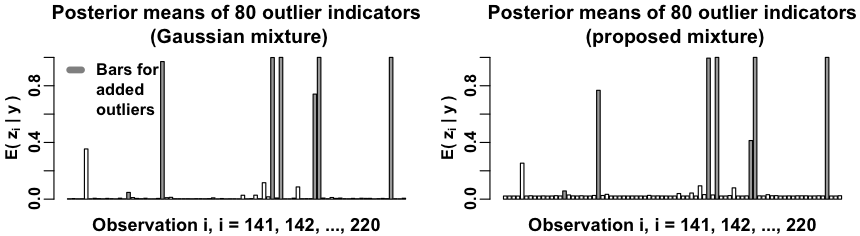}
\caption{Each panel shows posterior means of outlier indicators ($z_i$'s\textcolor{black}{, $i=141, 142, \ldots, 220$}) obtained by fitting the Gaussian mixture error model (left panel) and the proposed mixture error model (right panel) on $\boldsymbol{y}^{\textrm{sim}}$. The height of each bar represents the average of three million posterior samples of $z_i$ and the horizontal axis indicates observation $i$ ($i=\textcolor{black}{141, 142, \ldots, 220}$). \textcolor{black}{Both models clearly identify  six outliers, considering the gray bars.} Although the Gaussian mixture error model works better  than the proposed mixture error model in terms of designating non-inflated Gaussian errors to non-outlying observations correctly, it tends to designate inflated Gaussian errors to larger observations more often.}
\label{figure_macho_sim_indicator}
\end{centering}
\end{figure}
%

Bar plots in~Figure~\ref{figure_macho_sim_indicator} display the posterior means of \textcolor{black}{80} outlier indicators \textcolor{black}{from observation 141 to  220. (We choose this range to clarify seven gray bars corresponding to the seven added outliers.) In each panel, six gray bars are noticeably higher than the others, being flagged as outliers by both models, although the first gray bar for $y_{155}^{\textrm{sim}}$ is not; in the simulated data, more observations have the simulated brightness similar to $y_{155}^{\textrm{sim}}$, which makes $y_{155}^{\textrm{sim}}$ less likely to be an outlier. Most of the other} bars in the first panel have almost zero heights, while those in the second panel are slightly higher\textcolor{black}{; this pattern also appears for the data outside the range that are not displayed here.} This implies that the Gaussian mixture error model outperforms the proposed mixture error model in designating non-inflated Gaussian errors to non-outlying observations. However, when it comes to certain bars that are noticeably higher than the others, the Gaussian mixture error model tends to designate inflated Gaussian errors to them more often than the proposed mixture error model does (i.e., down-weighting larger observations \textcolor{black}{more often}). This makes the \textcolor{black}{former} produce a smaller estimate of $\log(\sigma)$ and a larger estimate of $\log(\tau)$ (due to negative association)  than the \textcolor{black}{latter} as shown in the second and third panels of Figure~\ref{figure_macho_sim}.

\begin{table}[t!]
\begin{center}\caption{Numerical summaries obtained by fitting the four error models on $\boldsymbol{y}^{\textrm{sim}}$; see the caption of Table~\ref{hospital_res_table} for the definitions of these summaries. As for the parameters of interest, i.e., $\log(\sigma)$ and $\log(\tau)$, the proposed mixture error model produces the most accurate estimates, considering that the MSE ratios are greater than 1, although it does not produce the shortest posterior interval for $\log(\sigma)$. Also, it takes  about 7\% more CPU time than the Gaussian mixture or $t_\nu$  error model. \textcolor{black}{We emphasize these aspects in bold font.}}\label{macho_sim_res_table}
{\small\begin{tabular}{ccrcrccr}
  & &  Posterior mean &&\multicolumn{1}{c}{MSE} & & Length & \multicolumn{1}{c}{CPU}\\
  & Error&\multicolumn{1}{c}{(Monte Carlo error)} & Bias  & \multicolumn{1}{c}{ratio} & \multicolumn{1}{c}{95\% P.I.} & of P.I. & \multicolumn{1}{c}{time}\\
\hline
& N  & 17.652 (0.00009)& 0.015  & 0.19 & (17.547, 17.764) & 0.217 &  470\\
\multirow{2}{*}{$\mu$}&$t_\nu$& 17.724 (0.00063)& 0.057 & 2.65 & (17.275, 18.299) & 1.024 & 504\\
&N+N& 17.709 (0.00058) & 0.042  & 1.44 & (17.360, 18.172) & 0.812 & 503\\
&N$+t_\nu$& 17.702 (0.00051) & 0.035  & - & (17.389, 18.120) & 0.731 & \textbf{540}\\\\
& N  & -3.303 (0.00033) & 0.715  &  \textbf{263.46} & (-3.471, -3.133) & 0.338 & 470\\
\multirow{2}{*}{$\log(\sigma)$}&$t_\nu$& -4.327 (0.00151) & 0.309 & \textbf{49.21} & (-4.581, -4.065) & 0.516 & 504\\
&N+N& -4.140 (0.00160) & 0.123  & \textbf{7.80} & (-4.382, -3.887) & 0.495 & 503\\
&N$+t_\nu$& -4.061 (0.00210) & \textbf{0.044}  & - & (-4.333, -3.797) & 0.536 & \textbf{540}\\\\
& N   & 4.227 (0.00146)& 1.422 & \textbf{24.88} & (3.681, 4.978) & 1.297 & 470\\
\multirow{2}{*}{$\log(\tau)$}&$t_\nu$ & 6.571 (0.00691) & 0.921  & \textbf{10.44} & (5.296, 9.123) & 3.827 & 504\\
&N+N & 6.115 (0.00708) & 0.466  & \textbf{2.67} & (4.973, 8.453) & 3.480 & 503\\
&N$+t_\nu$ & 5.934 (0.00731) & \textbf{0.285}  & - & (4.832, 8.175) & \textbf{3.343} & \textbf{540}\\
\end{tabular}}
\end{center}
\end{table}

Table~\ref{macho_sim_res_table} summarizes numerical results including the posterior mean, bias, MSE ratio, 95\% posterior interval and its length, and the CPU time in seconds; see the caption of Table~\ref{hospital_res_table} for details of their definitions.  As for the parameters of interest, $\sigma$ and $\tau$, the proposed mixture error model significantly improves estimation accuracy compared to the other error models, considering that \textcolor{black}{the biases are smaller than the others and} the MSE ratios are  greater than 1 \textcolor{black}{as emphasized in bold font}. Also, implementing the proposed mixture error model takes just about 7\% more CPU time  than running the Gaussian mixture  or $t_\nu$ error model. However, it turns out that the 95\% posterior interval for $\log(\sigma)$ obtained with the proposed mixture error model is not the shortest.

\subsubsection{Analysis of the observed data of MACHO  70.11469.82}\label{macho_real}

\begin{figure}[t!]
\begin{center}
\includegraphics[scale=0.425]{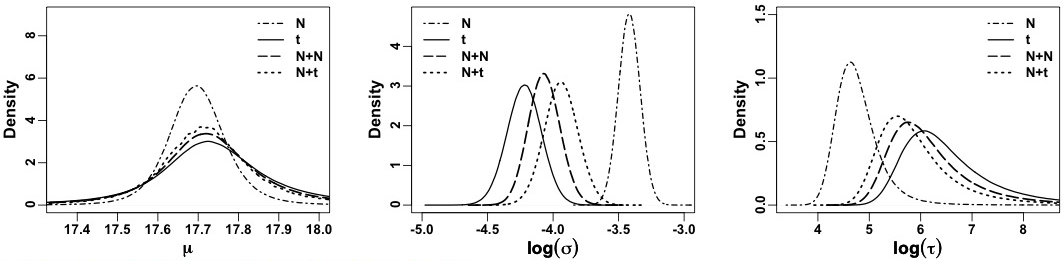}
\caption{The posterior distributions of $\mu$, $\log(\sigma)$, and $\log(\tau)$ (from the left panel) obtained by fitting Gaussian (dot-dashed), $t_\nu$ (solid), Gaussian mixture (dashed), and proposed mixture (dotted) error models on the observed data $\boldsymbol{y}$. These results are  almost identical to the outcomes of the simulation study displayed in Figure~\ref{figure_macho_sim}.}
\label{figure_macho}
\end{center}
\end{figure}

Finally, we fit the four error models on the data for MACHO  70.11469.82.  The sampling results are displayed in Figure~\ref{figure_macho} and are numerically summarized in Table~\ref{macho_res_table}.  These results are quite similar  to those of the simulation study shown and summarized in Figure~\ref{figure_macho_sim} and Table~\ref{macho_sim_res_table}, respectively. For example, the posterior distributions of $\log(\sigma)$ and $\log(\tau)$ from the Gaussian mixture error model in the second and third panel of Figure~\ref{figure_macho}, respectively, are located between those  from the $t_\nu$ error model and those from the proposed mixture error model, as is the case in Figure~\ref{figure_macho_sim}. \textcolor{black}{Also, Table~\ref{macho_res_table}  indicates that the proposed mixture error model produces the shortest  posterior interval for $\log(\tau)$ and takes slightly more CPU time than the other models as highlighted in bold font, which is the case in the simulation study.} Thus, even though we do not know the generative values for these MACHO time series data, it is likely that the proposed mixture error model might produce more accurate estimates than  the other error models for these data, analogous to the simulation study in Section~\ref{macho_sim}. Though not shown here, the result of outlier detection is also similar to that of  the simulation study displayed in Figure~\ref{figure_macho_sim_indicator}.



\begin{table}[t!]
\begin{center}\caption{Numerical summaries obtained by fitting the four error models on the observed data $\boldsymbol{y}$; see the caption of Table~\ref{hospital_res_table} for the computational details. Overall, these results are similar to those of the simulation study summarized in Table~\ref{macho_sim_res_table} \textcolor{black}{as highlighted in bold font}.}\label{macho_res_table}
{\small\begin{tabular}{ccrccr}
  & &  Posterior mean & & Length & \multicolumn{1}{c}{CPU}\\
  & Error&\multicolumn{1}{c}{(Monte Carlo error)}   & \multicolumn{1}{c}{95\% P.I.} & of P.I. & \multicolumn{1}{c}{time}\\
\hline
& N  & 17.699 (0.00019)&   (17.528, 17.882) & 0.354 &  467\\
\multirow{2}{*}{$\mu$}&$t_\nu$& 17.751 (0.00072)&  (17.241, 18.361) & 1.120 & 507\\
&N+N& 17.739 (0.00079) &  (17.309, 18.255) & 0.945 & 494\\
&N$+t_\nu$& 17.729 (0.00053) &  (17.359, 18.166) & 0.807 & \textbf{543}\\\\
& N  & -3.418 (0.00027) &  (-3.579, -3.254) & 0.325 & 467\\
\multirow{2}{*}{$\log(\sigma)$}&$t_\nu$& -4.220 (0.00118) &  (-4.481, -3.963) & 0.518 & 507\\
&N+N& -4.068 (0.00113)  & (-4.299, -3.829) & 0.470 & 494\\
&N$+t_\nu$& -3.939 (0.00203) & (-4.193, -3.683) & 0.510 & \textbf{543}\\\\
& N   & 4.785 (0.00260)& (4.134, 5.838) & 1.704 & 467\\
\multirow{2}{*}{$\log(\tau)$}&$t_\nu$ & 6.567 (0.00594) & (5.320, 9.105) & 3.785 & 507\\
&N+N & 6.215 (0.00499) &  (5.084, 8.575) & 3.491 & 494\\
&N$+t_\nu$ & 5.937 (0.00656) &  (4.879, 8.122) & \textbf{3.243} & \textbf{543}\\
\end{tabular}}\end{center}
\end{table}

\section{Concluding remarks}\label{sec:con}
A heavy tailed error assumption based on Student's $t$ distribution is well known for its robustness in parameter estimation compared to a commonly-used Gaussian error assumption. However, it may be inefficient to apply the heavy tailed error assumption to most of the data when majority of the errors are concentrated at zero. Thus we propose mixing the Gaussian and Student's $t$ errors by introducing latent outlier indicators, converting  Gaussian errors to $t$ errors only when the observed data are evaluated to be outliers. This mixture error assumption leverages the best of the Gaussian and $t$ error assumptions in that the resulting parameter estimation can be not only robust but also accurate. Using a Gaussian hierarchical model to fit the simulated hospital profiling data  and a state-space model of an Ornstein-Uhlenbeck process to fit the brightness time series data of a MACHO quasar, we have empirically shown that this mixture error can achieve both robustness and accuracy in estimating parameters.


There are several opportunities to build upon this work. First, we can extend the proposed mixture error in~\eqref{mixture_model} to even more general mixture errors by allowing any scale mixture family of a Gaussian distribution \citep{andrews1974scale, west1987scale}.  For example, if the prior distribution of $\alpha_i$ in~\eqref{mixture_model} is an Exponential($w^2$) distribution instead of the inverse-Gamma($\nu/2$, $\nu/2$) distribution, then the second mixture component in~\eqref{mix_marginal} becomes a Laplace($w$) distribution that is used for a Bayesian Lasso \citep{park2008lasso}. Second, \textcolor{black}{this mixture of Gaussian and Laplace distributions corresponds to Huber's loss function \citep{huber1964loss}, and thus it is meaningful to develop a non-Bayesian implementation of the proposed mixture of Gaussian and $t$ distributions to compare these two mixtures under Huber's framework. Third,} converting  Gaussian errors into  mixture errors can be simply achieved  as illustrated, but it is unclear whether the conversion automatically guarantees posterior propriety when the original Gaussian error model guarantees it with jointly improper  prior distributions.   Another avenue for further improvement is to derive an optimization-based inference for a mixture error model using an EM algorithm \citep{dempster1977maximum} as is usually done for mixture models \citep{aitkin1980mixture}. Finally, for some cases it is desirable to consider the measurement covariance matrix $\boldsymbol{V}_i$ in~\eqref{mixture_model} as unknown. We invite interested readers to explore these possibilities.

\bigskip
\noindent{\Large\bf Supplementary materials}

\begin{description}

\item[Appendices:] Appendices A, B, and C cited in the article (Appendices.pdf).

\item[R code and data:] All of the R codes and data used in this article (RcodeData.zip).

\end{description}

\bigskip
\noindent{\Large\bf Acknowledgements}\\
Hyungsuk Tak and Sujit Ghosh acknowledge partial support from the NSF grant DMS 1127914 (and DMS 1638521 only for Hyungsuk Tak) given to the Statistical and Applied Mathematical Sciences Institute.  Justin Ellis acknowledges support by NASA through Einstein Fellowship grant PF4-150120. We also thank Xiao-Li Meng and David van Dyk for  helpful discussions, the editor, associate editor, and two referees for insightful comments and suggestions, and Steven Finch for his careful proofreading.

\bibliography{bibliography}
\bibliographystyle{apalike}

\begin{appendices}

\section{\!\!\!\!\!.~ The marginal posterior density and posterior propriety in Section~1}\label{app1}

With the improper flat prior distribution (Lebesgue) on $\mu$, the posterior distribution of $\mu$ based on the Gaussian error is N$_1(\bar{y},~1/20)$, where $\bar{y}$ is the sample mean of the data. Clearly, this posterior density is proper.

The full posterior distribution based on the $t_4$ error model is 
\begin{equation}\label{full_posterior_app1}
\pi_1(\mu, \boldsymbol{\alpha}\mid \boldsymbol{y})\propto q(\mu) h(\boldsymbol{\alpha}) \prod_{i=1}^{20}\textrm{N}_1(y_i\mid \mu,~ \alpha_iV_i),
\end{equation}
where $q(\mu)\propto 1$ and $h(\boldsymbol{\alpha})$ is proportional to the product of  inverse-Gamma($\nu/2, \nu/2$) prior densities of $\alpha_i$'s. With $\nu=4$, the marginal posterior density of $\mu$ with $\boldsymbol{\alpha}$ integrated out from~\eqref{full_posterior_app1} is 
\begin{equation}\label{marginal_posterior_app1}
\pi_2(\mu\mid \boldsymbol{y})\propto \prod_{i=1}^{20}(1+(y_i-\mu)^2/4)^{-2.5},
\end{equation}
where the right-hand side is the product of the densities of a shifted $t_4$-distribution.  This  posterior density of $\mu$ is proper because  an upper bound of~\eqref{marginal_posterior_app1}, i.e., $(1+(y_1-\mu)^2/4)^{-2.5}$, results in a finite integral with respect to $\mu$. Thus the joint posterior in~\eqref{full_posterior_app1} is also proper. 

The full posterior distribution based on the mixture error model is
\begin{equation}\label{full_posterior2_app1}
\pi^\ast_1(\mu, \boldsymbol{\alpha}, \boldsymbol{z}\mid \boldsymbol{y})\propto q(\mu) h(\boldsymbol{\alpha}) p(\boldsymbol{z}) \prod_{i=1}^{20}\textrm{N}_1(y_i\mid \mu, ~\alpha_i^{z_i}V_i),
\end{equation}
where $q$ and $h$ are the same density functions used in~\eqref{full_posterior_app1}, and $p$ is proportional to the product of  Bernoulli(0.1) prior mass functions of $z_i$'s. With $\nu=4$, the  posterior density of $\mu$ and $\boldsymbol{\alpha}$ with $\boldsymbol{z}$ integrated out from~\eqref{full_posterior2_app1} is
\begin{align}\label{marginal_posterior2_app1}
\begin{aligned}
\pi^\ast_2(\mu, \boldsymbol{\alpha}\mid \boldsymbol{y}) \propto &\prod_{i=1}^{20}\left[ 0.1\times\alpha_i^{-0.5}\exp(-(y_i-\mu)^2/(2\alpha_i))+0.9\times\exp(-(y_i-\mu)^2/2))\right]\\
&\times \prod_{i=1}^{20}\alpha_i^{-3}\exp(-2/\alpha_i).
\end{aligned}
\end{align}
The marginal posterior density of $\mu$ with $\boldsymbol{\alpha}$ integrated out from~\eqref{marginal_posterior2_app1} is
\begin{equation}\label{marginal_posterior3_app1}
\pi^\ast_3(\mu\mid \boldsymbol{y}) \propto \prod_{i=1}^{20}\left[ 0.1\times(1+(y_i-\mu)^2/4)^{-2.5}+0.9\times\exp(-(y_i-\mu)^2/2))\right],
\end{equation}
whose tails decay as a power law, $(1+\vert\mu\vert)^{-100}$, and thus the integral of $\pi^\ast_3(\mu\mid \boldsymbol{y})$ with respect to $\mu$ is finite. Consequently, the full posterior distribution in~\eqref{full_posterior2_app1} is proper. 



\section{\!\!\!\!\!.~ Details in Section~3.1}\label{app2}

\subsection{The Gibbs sampler}\label{app2se1}
To sample the full posterior distribution in (8) that is based on a Gaussian error assumption, we derive a Gibbs sampler that iteratively samples the  three conditional posterior distributions outlined in~(9), i.e., for $i=1, \ldots, 31$,
\begin{align}\label{gibbs_hospital}
\begin{aligned}
\mu_i \mid \beta, A, \boldsymbol{y} &\sim \textrm{N}_1\!\left((1-B_i)y_i,~ (1-B_i)V_i\right),\\
\beta \mid \boldsymbol{\mu}, A, \boldsymbol{y} &\sim \textrm{N}_1\!\left(\frac{(31/A)\bar{\mu}}{(31/A)+(1/10^5)},~\frac{1}{(31/A)+(1/10^5)}\right)\!,\\
\pi_3(A \mid \beta, \boldsymbol{\mu}, \boldsymbol{y}) &\propto (10^5+A)^{-2}\times  \prod_{i=1}^{31}\textrm{N}_1(\mu_i\mid \beta, A) ,\\
\end{aligned}
\end{align}
where $B_i=V_i / (V_i + A)$ is a shrinkage factor and $\bar{\mu}$ is the sample mean of $\boldsymbol{\mu}$. Since the conditional posterior distribution of $A$ cannot be sampled directly, we use a Metropolis-Hastings algorithm to sample $A$ within the Gibbs sampler (Tierney, 1994). We draw a proposal $\log(A^\ast)$ from N$_1(\log(A^{(i-1)})\mid \sigma^2)$ at iteration $i$, where the proposal scale $\sigma$ is adaptively set to produce the acceptance rate around 0.35 for all of the error models in each case. We set $A^{(i)}$ to $A^\ast$ with a probability 
\begin{equation}\label{accept_rate_uni}
\min\left[1,~\frac{p(A^\ast\mid \boldsymbol{\beta}^{(i)},  \boldsymbol{\mu}^{(i)}, \boldsymbol{y})}{p(A^{(i-1)}\mid \boldsymbol{\beta}^{(i)},  \boldsymbol{\mu}^{(i)}, \boldsymbol{y})}\times \frac{A^\ast}{A^{(i-1)}}\right]
\end{equation}
and set $A^{(i)}$ to $A^{(i-1)}$ otherwise. The  ratio $A^\ast/A^{(i-1)}$ in~\eqref{accept_rate_uni} is the Hastings ratio for the update of $A$ on a logarithmic scale.

The extended full posterior distribution based on a mixture error assumption is specified in~(11). An extended Gibbs sampler uses the conditional posterior distributions of the original Gibbs sampler in~\eqref{gibbs_hospital}  to sample $\boldsymbol{\mu}$, $\beta$, and $A$ after  replacing $V_i$ (including those in $B_i$) with $\alpha^{z_i}_iV_i$ in the conditional posterior distribution of $\mu_i$. After updating $\boldsymbol{\mu}$, $\beta$, and $A$, the extended Gibbs sampler updates the additional parameters, i.e., $\boldsymbol{z}$ and $\boldsymbol{\alpha}$ via~(12) and $\theta$ and $\nu$ via~(5). As for the initial values of this extended Gibbs sampler, we set $\mu^{(0)}_i=y_i^{\textrm{sim}}$, $A^{(0)}=\sum_{i=1}^{31}V_i/31$, $\beta^{(0)}=\bar{y}^{\textrm{sim}}$, $z_i^{(0)} = 0$ ($z_i^{(0)} = 1$ only for the $t_\nu$ error model), $\alpha_i^{(0)}=1$, $\theta^{(0)}=0.01$ for all~$i$. We use this extended Gibbs sampler to obtain  sampling results for all of the error models; see Section~3.1.1 for details.
%

The Gaussian mixture error model  assumes that $\alpha_i=\alpha$ for all $i$. Based on this assumption, the marginalized likelihood function for $\beta$, $\theta$, $A$, and $\alpha$ is
\begin{equation}\label{jointlik}
L(\beta, \theta, A, \alpha)\propto \prod_{i=1}^{31}\left[\theta \textrm{N}_1(y_i\mid \beta,~ A+\alpha V_i)+(1-\theta)\textrm{N}_1(y_i\mid \beta,~ A+V_i)\right].
\end{equation}
We obtain the maximum likelihood estimates,  $\hat{\beta}, \hat{\theta}, \hat{A}$, and $\hat{\alpha}$,  that jointly maximize~\eqref{jointlik}. To obtain the sampling result of the Gaussian mixture error model, we set $\alpha_i^{(0)}=\hat{\alpha}$ for all~$i$ in the extended Gibbs sampler without updating $\boldsymbol{\alpha}$ and $\nu$.


\subsection{Sensitivity analyses according to $k$, $m$, and the data generation assumption}\label{app2sec2}

Using the simulated data with synthetic outliers, $\boldsymbol{y}^{\textrm{out}}$, we conduct a sensitivity analysis for the posterior inference on $\log(A)$  of the proposed mixture error model according to various Beta($km$, $k(1-m)$) prior distributions of $\theta$. The posterior inference on $\beta$ does not reveal noticeable differences as is the case in the top-right panel of Figure~2.

\begin{figure}[t!]
\begin{center}
\includegraphics[scale=0.425]{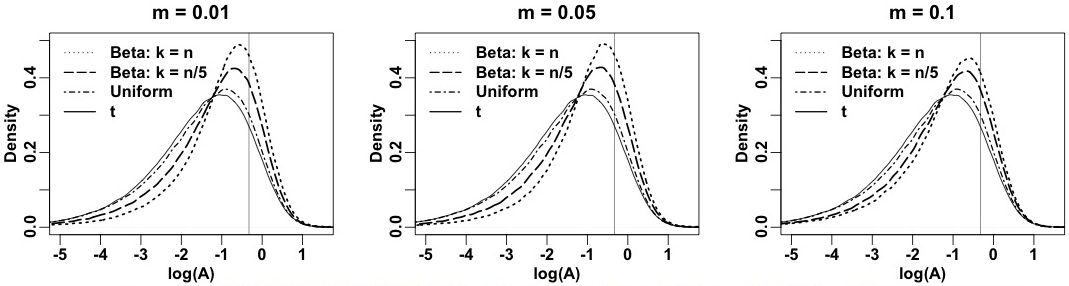}
\caption{The results of sensitivity analysis for the inference on $\log(A)$. We fit the proposed mixture error model on $\boldsymbol{y}^{\textrm{out}}$ with various Beta($km, k(1-m)$) prior distributions on $\theta$. Each panel shows the posterior densities of $\log(A)$ obtained by the $t$ error model and  proposed mixture error model with three different priors on $\theta$. The three panels show the results obtained with three different values of $m$. \textcolor{black}{The vertical lines indicate the generative true values.} These sensitivity analyses indicate that the resulting inference of the mixture error model becomes close to that of the $t_\nu$ error model as the Beta prior approaches the Uniform(0, 1) prior, i.e., as $k$ decreases and $m$ increases.}
\label{figure_app2}
\end{center}
\end{figure}

Figure~\ref{figure_app2} displays the posterior densities of $\log(A)$.  In each panel,  we denote the posterior density obtained by the $t_\nu$ error model by the solid curve to compare it with other posterior densities. The posterior density obtained with a strong Beta prior ($k=n$) is denoted by the dotted curve, that with a weak Beta prior ($k=n/5$) is represented by the dashed curve, and that with a Uniform(0, 1) prior is denoted by the dot-dashed curve. The three panels show the results with three different values of $m$, i.e., 0.01, 0.05, and 0.1.  Clearly, the resulting inference on $\log(A)$ obtained by the proposed mixture error model is sensitive to the choices of $k$ and $m$. The inference becomes close to the one obtained by the $t_\nu$ error model as $k$ decreases and $m$ increases (to 0.5), i.e., as the Beta prior moves towards the Uniform(0, 1). 

\begin{figure}[t!]
\begin{center}
\includegraphics[scale=0.425]{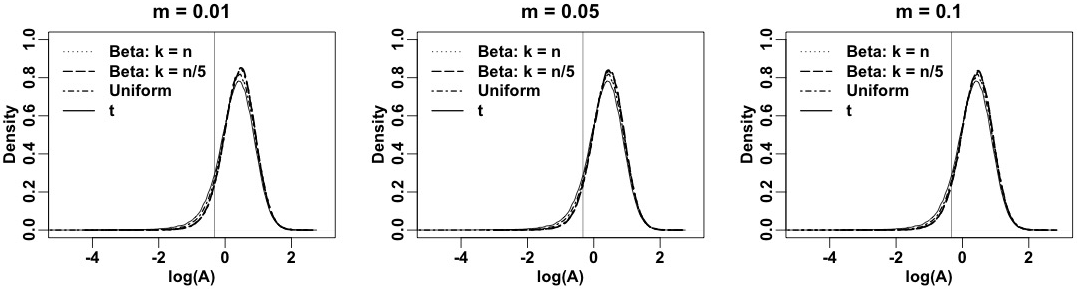}
\caption{The results of sensitivity analysis when the simulated data are generated with $t_4$ errors; we do not introduce synthetic outliers. The format of each panel is the same as the one in Figure~\ref{figure_app2}. The inference obtained with each mixture error model is close to the one with the $t_\nu$ error model, considering that the data are simulated with $t_4$ errors.}
\label{figure_app2_4}
\end{center}
\end{figure}

We also conduct another sensitivity analysis to see the impact of the data generation assumption. This time we newly simulate a data set with $t_4$ errors instead of Gaussian errors; we sample $\boldsymbol{\mu}^{\textrm{sim}}$ given $\beta_{\textrm{gen}}=0$ and $A_{\textrm{gen}}=0.722$ using~(6), and then  independently generate $y^{\textrm{sim}}_i$  using a $\mu^{\textrm{sim}}_i+ V_i^{0.5}t_4$ distribution, where $\mu^{\textrm{sim}}_i$ is the location parameter and $V_i^{0.5}$ is the scale parameter of the $t_4$ distribution. We do not introduce synthetic outliers. Figure~\ref{figure_app2_4} displays the posterior densities of $\log(A)$ obtained by the $t_\nu$ error model and proposed mixture error model with different Beta priors on~$\theta$ in the same format as Figure~\ref{figure_app2}. It shows that the posterior densities of $\log(A)$ obtained by both $t_\nu$ and mixture error models are close to each other, although that obtained by $t_\nu$ error model puts slightly more mass near $\log(A_{\textrm{gen}})$.

\subsection{Sensitivity analyses according to the data size and outlier proportion}\label{app2sec3}

Here we conduct sensitivity analyses to see the impact of  data size and proportion of outliers on the posterior inference. We consider nine cases as combinations of three data sizes, 20, 50, and 100, and three outlier proportions, 10\%, 20\% and 30\%. This time we assume that $V_i=1$, $\beta_{\textrm{gen}}=0$, and $A_{\textrm{gen}}=1$. Given the generative values, $\beta_{\textrm{gen}}$ and $A_{\textrm{gen}}$, we generate 100 simulated data points, $\boldsymbol{y}^{\textrm{sim}}\equiv\{y_1^{\textrm{sim}}, y_2^{\textrm{sim}}, \ldots, y_{100}^{\textrm{sim}}\}$, using the marginalized sampling distribution of $y_i$, i.e., N$_1(\beta_{\textrm{gen}}, 1 + A_{\textrm{gen}})$ with $\mu_i$ integrated out. These data  points are the same for all of the nine cases. For the cases with $n=20$ (or $n=50$), we use the first 20 (or 50) values of $\boldsymbol{y}^{\textrm{sim}}$. To generate synthetic outliers, we generate outliers from N$_1(0, 20^2)$ according to the designated proportions, and replace the simulated data with these outliers. For the case of $n=20$ and 10\% outliers, for example, the data set is composed of the first 20 values of $\boldsymbol{y}^{\textrm{sim}}$, and we replace its first two values  with  two synthetic outliers generated from N$_1(0, 20^2)$.

\begin{figure}[t!]
\begin{center}
\includegraphics[scale=0.425]{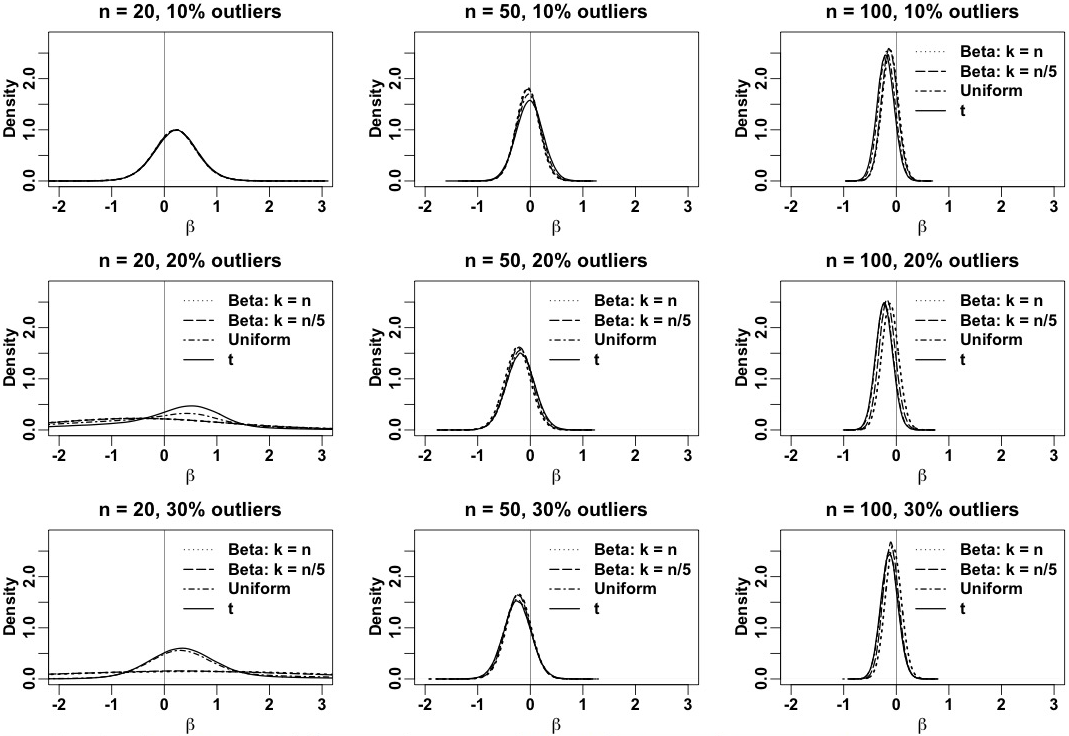}
\caption{The result of sensitivity analysis for the posterior density of $\beta$ according to the data size and outlier proportion. Each panel shows four density curves obtained by different error models. \textcolor{black}{The vertical lines indicate the generative true values.} It shows that a weak prior on $\theta$, e.g., Uniform(0, 1), can prevent a misleading inference when the date size is small and outlier proportion is large; see the second and third panels in the first column.}
\label{figure_app6}
\end{center}
\end{figure}

\begin{figure}[t!]
\begin{center}
\includegraphics[scale=0.425]{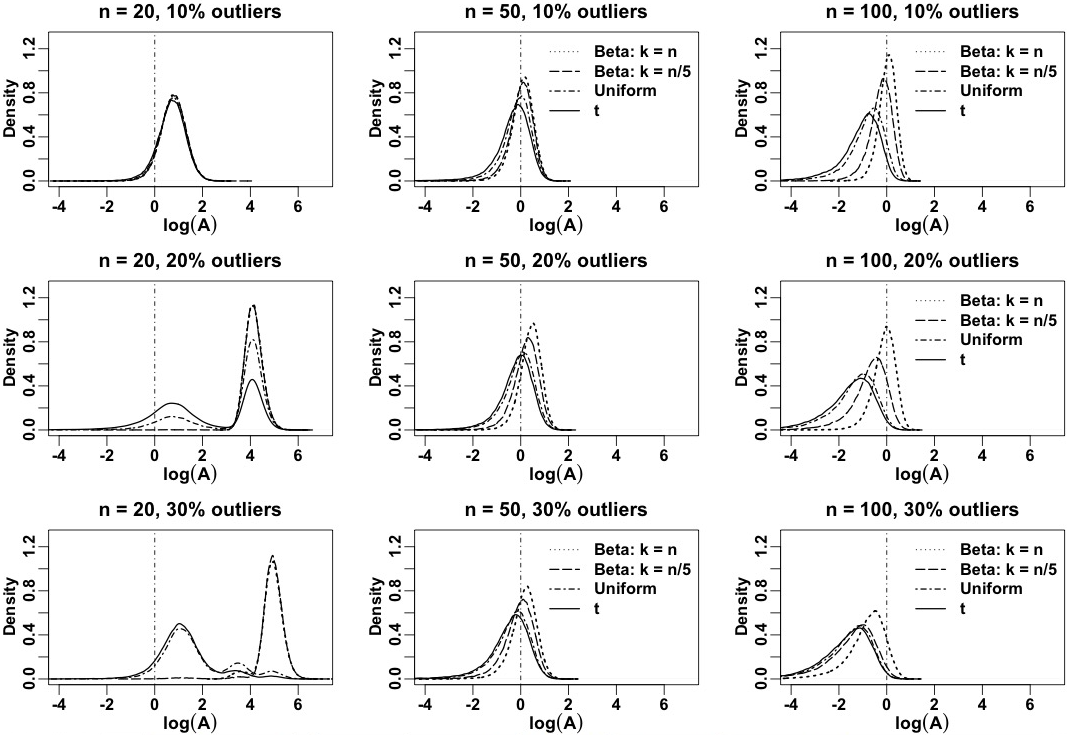}
\caption{The result of sensitivity analysis for the posterior density of $\log(A)$ according to the data size and outlier proportion. Each panel shows four density curves obtained by different error models. \textcolor{black}{The vertical dot-dashed lines indicate the generative true values.}  It shows that a weak prior on $\theta$, e.g., Uniform(0, 1), can prevent a misleading inference when the date size is small; see the second and third panels in the first column.}
\label{figure_app7}
\end{center}
\end{figure}

We fit both $t_\nu$ and mixture error models on the nine data sets with three different Beta priors on $\theta$  for the mixture model as is the case in Section~\ref{app2sec2}; we  consider $m=0.01$ if applicable. For each model and case, we run a single Markov chain with length 550,000 and discard the first 50,000 as burn-in. We summarize the sampling result of $\beta$ in Figure~\ref{figure_app6} and that of $\log(A)$ in Figure~\ref{figure_app7}. The mixture error model with $k=n$ or $k=n/5$ performs poorly under the cases where $n=20$ with large proportions of outliers ($20\%, 30\%$). It results in an extremely wide spread for the density of $\beta$ and severe bias for the density of $\log(A)$; see the second and third panels in the first column of Figure~\ref{figure_app6} and those of Figure~\ref{figure_app7}.  \textcolor{black}{We} notice that this result is similar to the inference of the Gaussian error model in the presence of outliers in Section~3.1.2. This happens because the Beta($km, k(1-m)$) prior with $k=n$ (or $k=n/5$) and $m=0.01$ is strong enough  to designate Gaussian errors to outlying observations a posteriori, making the resulting inference similar to that obtained with  Gaussian errors\footnote{\textcolor{black}{In Figure 12, the estimation accuracy improves in the third panel of the first column compared to that in the second panel. The reason is that the randomly generated outliers from N(0, $20^2$)  are quite different between two cases; four data points are generated to be outliers in the second panel and six data points in the third panel. The biggest  outlier in the second panel is 20.6 and that in the third panel is $-40.6$. It makes the two mixture models with strong Beta priors ($k=n$ and $k=n/5$) produce even larger posterior samples of the variance component, $A$, in the third panel because such strong Beta priors let the two mixture error model behave similarly to the Gaussian error model. On the other hand, two out of the six added data points in the third panel are by chance centrally located, and also the other three are less severe outliers than those in the second panel. These allow the $t$ error model and the mixture error model with a weak Beta prior (Uniform) to produce smaller posterior samples of $A$ concentrating more on the generative value of $A$ in the third panel.}}.  These results indicate that a weak prior on $\theta$, e.g., Uniform(0, 1),  is desirable and safe when the data size is small. In other cases, the strong Beta prior with $k=n$ (or $k=n/5$) tend to produce more accurate inference.

\subsection{MCMC convergence diagnostics}\label{app2sec4}

We check the convergence of the Markov chain that was used in Section~3.1.2. For the posterior inference, we independently implemented 30 Markov chains each for 1,050,000 iterations and discarded the first 50,000  as burn-in iterations. We thinned each chain from length 1,000,000 to 100,000 and combined the 30 thinned Markov chains. Thus, the length of the combined Markov chain is 3,000,000.

\begin{figure}[t!]
\begin{center}
\includegraphics[scale=0.425]{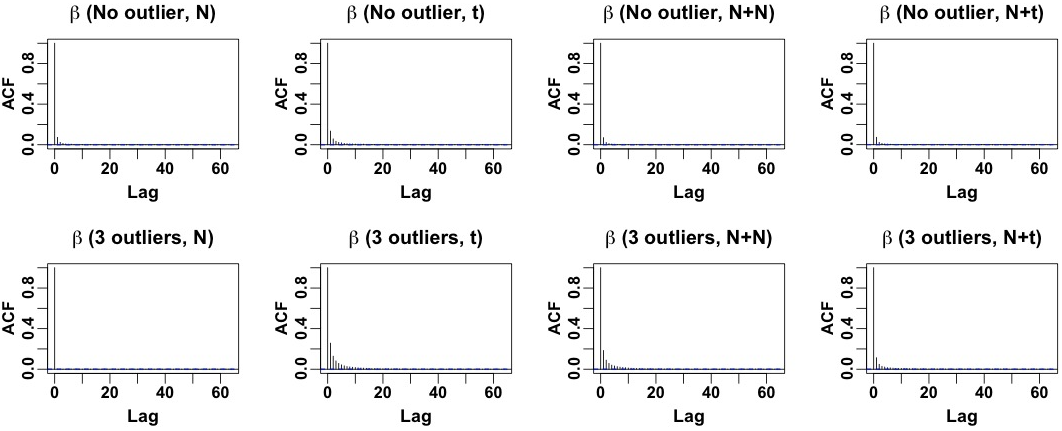}
\caption{The auto-correlation functions of $\beta$ obtained by four different error models, i.e., Gaussian (N), $t_\nu$ (t), Gaussian mixture (N+N), and proposed mixture (N+t) error models. The length of (thinned and combined) Markov chain is 3,000,000. The upper panels show the case without an outlier and the bottom panels exhibit the case with three synthetic outliers. These auto-correlation functions decrease quickly, and thus the Markov chain convergence appears satisfactory.}
\label{figure_app2_2}
\end{center}
\end{figure}

The first row of Figure~\ref{figure_app2_2} shows four auto-correlation functions of $\beta$ obtained by four different error models under the case without an outlier. \textcolor{black}{The effective sample sizes\footnote{\textcolor{black}{We use a function \texttt{effectiveSize} of an R package \texttt{coda} \citep{plummer2006coda} to estimate the effective sample size.}} (ESSs) of each combined posterior sample of $\beta$ divided by the total number of iterations, i.e., ESSs per iteration, are 0.790, 0.621, 0.799, and 0.770 for the Gaussian, $t_\nu$, Gaussian mixture, and proposed mixture error models, respectively. Also, the ESSs divided by the CPU times (seconds), i.e., ESSs per second, are 84607, 35164, 49965, and 31217, for the four error models, respectively. } Both auto-correlation function and \textcolor{black}{ESS} do not indicate any lack of convergence. Similarly, the second row displays those under the case with three synthetic outliers. The \textcolor{black}{ESSs per iteration} are 1.000, 0.392, 0.476, and 0.586, and those per \textcolor{black}{second} are 107143, 22173, 29745, and 23778 \textcolor{black}{for the Gaussian, $t_\nu$, Gaussian mixture, and proposed mixture error models, respectively}. All of the auto-correlation functions decrease quickly and the \textcolor{black}{ESSs} are large without showing any evidence of the lack of convergence.

\begin{figure}[b!]
\begin{center}
\includegraphics[scale=0.425]{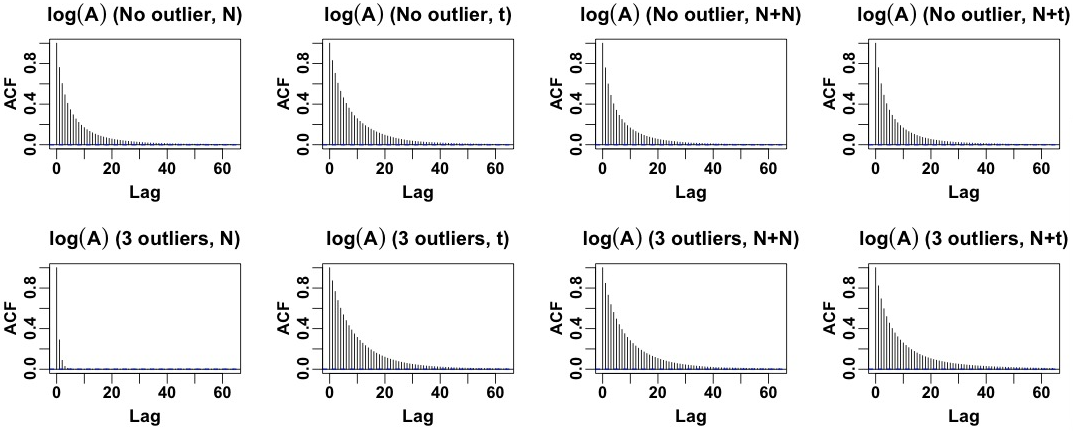}
\caption{The auto-correlation functions of $\log(A)$ obtained by four different error models, i.e., Gaussian (N), $t_\nu$ (t), Gaussian mixture (N+N), and proposed mixture (N+t) error models. The length of (thinned and combined) Markov chain is 3,000,000. The upper panels show the case without an outlier and the bottom panels exhibit the case with three synthetic outliers. These auto-correlation functions do not indicate the lack of convergence.}
\label{figure_app2_2_2}
\end{center}
\end{figure}

In Figure~\ref{figure_app2_2_2}, we display the auto-correlation functions of $\log(A)$ in the same format as Figure~\ref{figure_app2_2}. The auto-correlation functions decrease quickly in all of the cases. \textcolor{black}{When there is no outlier, the ESSs per iteration of each combined posterior sample of $\log(A)$ corresponding to the first row of Figure~\ref{figure_app2_2_2} are 0.212, 0.132, 0.210, and 0.208 for the Gaussian, $t_\nu$, Gaussian mixture, and proposed mixture error models, respectively. Also, their ESSs per seconds are 22739, 7446, 13139, and 8450, respectively. When there exist three  outliers, the ESSs per iteration are 0.601, 0.124, 0.143, and 0.030, and the ESSs per second are 64429, 7015, 8924, and 1199 for the four models, respectively}. The \textcolor{black}{ESS} of the Gaussian error model is striking, though the resulting inference is severely biased as shown in the bottom-right panel of Figure~3. Although the \textcolor{black}{ESS} of the proposed mixture error model is smaller than the others, the Markov chain convergence might not be a serious issue  here because the auto-correlation functions are similar to each other, decreasing quickly.

\begin{figure}[t!]
\begin{center}
\includegraphics[scale=0.425]{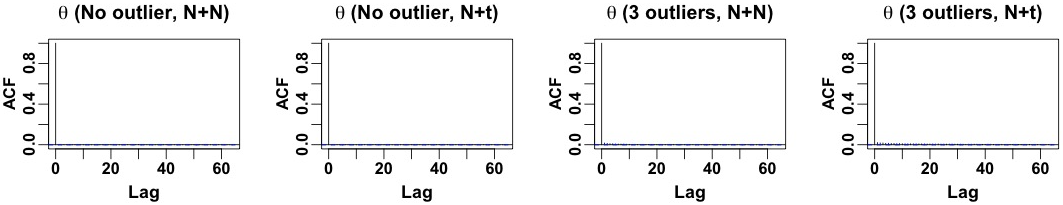}
\caption{The auto-correlation functions of $\theta$ obtained by Gaussian (N+N)  and proposed (N+t) mixture error models. The length of (thinned and combined) Markov chain is 3,000,000. The first two panels show the case without an outlier and the last two panels exhibit the case with three synthetic outliers. These functions decrease to zero immediately.}
\label{figure_app2_3}
\end{center}
\end{figure}

Finally, instead of checking the convergence of each outlier indicator, we check the auto-correlation function and \textcolor{black}{ESS} of $\theta$ obtained by the Gaussian and proposed mixture error models because $\theta$ governs the outlier indicators. Figure~\ref{figure_app2_3} displays the auto-correlation functions of $\theta$ obtained by the two mixture error models under two different cases; no outlier and three synthetic outliers. These auto-correlation functions immediately decrease to zero. The \textcolor{black}{ESSs per iteration} are 0.998, 0.992, 0.830, and 0.563 from the left, and \textcolor{black}{the ESSs per second} are 106884, 56175, 51858, and 22836. Although the \textcolor{black}{ESS} of the proposed mixture error model is the smallest, its auto-correlation function is almost zero from the beginning, and thus we do not consider the smallest \textcolor{black}{ESS} as the evidence of the lack of convergence.

\section{\!\!\!\!\!.~ Details in Section~3.2}\label{app3}

\subsection{The Gibbs sampler}\label{app3sec1}

We use  a Metropolis-Hastings within Gibbs sampler to sample the full posterior distribution in~(17) that is based on the Gaussian error assumption, iteratively sampling the following conditional posterior distributions (also mentioned in~(18)):
\begin{align}\label{gibbs_app2}
\begin{aligned}
\pi_1(\boldsymbol{Y}(\boldsymbol{t})\mid \mu, \sigma^2, \tau, \boldsymbol{y}),~~\pi_2(\mu\mid \boldsymbol{Y}(\boldsymbol{t}), \sigma^2, \tau, \boldsymbol{y}),\\
\pi_3(\sigma^2 \mid \mu, \boldsymbol{Y}(\boldsymbol{t}), \tau, \boldsymbol{y}),~~\pi_4(\tau \mid \sigma^2, \mu, \boldsymbol{Y}(\boldsymbol{t}), \boldsymbol{y}).
\end{aligned}
\end{align}
The  conditional posterior distribution of $\tau$ is not a standard family distribution while the others can be directly sampled. Thus we adaptively sample $\tau$ via a Metropolis-Hastings kernel whose invariant distribution is $\pi_4$ in~\eqref{gibbs_app2}; see Appendices~\ref{subapp1} and \ref{subapp2} below for details of~\eqref{gibbs_app2}.

The extended full posterior distribution based on the mixture error assumption is specified in~(20). A corresponding extended Gibbs sampler  uses the conditional posterior distributions of the original Gibbs sampler in~\eqref{gibbs_app2} to update $\boldsymbol{Y}(\boldsymbol{t})$, $\mu$, $\sigma^2$, and $\tau$ after replacing $V_i$  in $\pi_1(\boldsymbol{Y}(\boldsymbol{t})\mid \mu, \sigma^2, \tau, \boldsymbol{y})$ with $\alpha_i^{z_i}V_i$. After updating these parameters, the extended Gibbs sampler updates the additional parameters, i.e., $\boldsymbol{z}$ and $\boldsymbol{\alpha}$ via~(21) and $\theta$ and $\nu$ via~(5). 

The initial values for Markov chains of each error model are  $Y^{(0)}(t_i)=y_i$, $\mu^{(0)} = \bar{y}$, $\sigma^{(0)}=0.01$, $\tau^{(0)}=200$,  $z_i^{(0)} = 0$ ($z_i^{(0)} = 1$ only for the $t$ error model), $\alpha_i^{(0)}=1$, $\theta^{(0)}=0.01$ for all~$i$. For the Gaussian mixture error model, we set $\alpha^{(0)}_i = 10^2$ for all $i$ and do not update $\alpha_i$'s and $\nu$ during the run.

\subsubsection{Conditional posterior distributions of $\boldsymbol{Y(t)}$}\label{subapp1}

We define $y'_{i}\equiv y_i-\mu$ and $Y'(t_i)\equiv Y(t_i) - \mu$. Let ``$< t_i$'' denote a set $\{t_k:~k=1, 2, \ldots, i-1\}$, ``$> t_i$'' denote $\{t_k:~k=i+1, i+2, \ldots, n\}$, and $a_i=\exp(-(t_i-t_{i-1})/\tau)$ for $i=2, 3, \ldots, 242$. To sample $\pi_1(\boldsymbol{Y}(\boldsymbol{t})\mid \mu, \sigma^2, \tau, \boldsymbol{y})$ in~\eqref{gibbs_app2}, we sample the following conditional posterior distributions. We suppress conditioning on $\mu, \sigma^2, \tau$, and $\boldsymbol{y}$ to save space.
{\begin{equation}
Y'(t_1) \mid \boldsymbol{Y}'(>t_1)\sim \textrm{N}_1\left[(1-B_1)y'_{1}+B_1a_2Y'(t_2), ~(1-B_1)V_{1}\right],
\end{equation}
where $B_1=V_{1}~/~[V_{1}+\tau\sigma^2(1-a_2^2)/2]$. For $i=2, 3, \ldots, 241$,
\begin{align}
\begin{aligned}
&Y'(t_i) \mid \boldsymbol{Y}'(<t_i), \boldsymbol{Y}'(>t_i)\\
&\sim\! \textrm{N}_1\!\left[(1-B_i)y'_{i}+B_i\left((1-B_i^\ast)\frac{Y'(t_{i+1})}{a_{i+1}}+B_i^\ast a_iY'(t_{i-1})\right),~ (1-B_i)\alpha^{z_i}V_{i}\right]\!,
\end{aligned}
\end{align}
where 
\begin{displaymath}
B_i=\frac{V_{i}}{V_{i}+\frac{\tau\sigma^2}{2}\frac{(1-a^2_{i})(1-a^2_{i+1})}{1-a^2_{i}a^2_{i+1}}}~~\textrm{and}~~B_i^\ast=\frac{1-a^2_{i+1}}{1-a^2_i a^2_{i+1}}.
\end{displaymath} 
Lastly,
\begin{equation}
Y'(t_{242}) \mid \boldsymbol{Y}'(<t_{242})\sim \textrm{N}_1\left[(1-B_{242})y'_{242} + B_{242}a_{242}Y'(t_{245}),~ (1-B_{242})\alpha^{z_{242}}V_{242}\right],
\end{equation}
where $B_{242}=V_{242}/[V_{242}+\tau\sigma^2(1-a_{242}^2)/2]$. }

\subsubsection{Conditional posterior distributions of the O-U parameters}\label{subapp2}

We use the same notation $Y'(t_i)$ and $a_i$ as in Appendix \ref{subapp1}. We sample $\pi_2$ in~\eqref{gibbs_app2} using a truncated Gaussian posterior distribution whose support is $(-30,~ 30)$:
\begin{equation}\label{ou_mu}
\mu \mid \boldsymbol{Y}(\boldsymbol{t}), \sigma^2, \tau, \boldsymbol{y}\sim\textrm{N}_1\!\left[\frac{Y(t_1) + \sum_{i=2}^{242} \frac{Y(t_i) - a_i Y(t_{i-1})}{1 + a_i}}{1 + \sum_{i=2}^{242} \frac{1-a_i}{1 + a_i}},~\frac{\tau\sigma^2/2}{1 + \sum_{i=2}^{242} \frac{1-a_i}{1 + a_i}}\right]\!.\nonumber
\end{equation}
We sample $\pi_3(\sigma^2 \mid \mu, \boldsymbol{Y}(\boldsymbol{t}), \tau, \boldsymbol{y})$ in~\eqref{gibbs_app2} using the following inverse-Gamma distribution:
\begin{equation}
\sigma^2 \mid \mu, \boldsymbol{Y}(\boldsymbol{t}), \tau, \boldsymbol{y}\sim \textrm{inverse-Gamma}\!\left(\frac{n+2}{2},~ 10^{-7}+\frac{Y'(t_1)^2}{\tau}+\sum_{i=2}^{242} \frac{\big[Y'(t_i) - a_iY'(t_{i-1})\big]^2}{\tau(1-a_i^2)}\right)\!.\nonumber
\end{equation}
Finally, we use a Metropolis-Hastings algorithm to sample 
\begin{equation}
\pi_4(\tau \mid \sigma^2, \mu, \boldsymbol{Y}(\boldsymbol{t}), \boldsymbol{y})\propto\frac{\exp\left(-\frac{1}{\tau}-\frac{Y'(t_1)^2}{\tau\sigma^2}-\sum_{i=2}^{242}\frac{\big[Y'(t_i) - a_iY'(t_{i-1})\big]^2}{\tau\sigma^2(1-a^2_i)}\right)I_{\{\tau>0\}}}{\tau^{(242+4)/2}\prod_{i=2}^{242}(1-a_i^2)^{0.5}}.\nonumber
\end{equation}
At iteration $i$, we draw a proposal $\log(\tau^\ast)$ from N$_1(\log(\tau^{(i-1)}), \phi^2)$, where $\phi$ is the proposal scale.  We set $\tau^{(i)}$ to $\tau^\ast$ with a probability 
\begin{equation}\label{accept_rate_uni2}
\min\left[1,~\frac{\pi_4(\tau^\ast \mid (\sigma^2)^{(i)}, \mu^{(i)}, \boldsymbol{Y}(\boldsymbol{t}), \boldsymbol{y})}{\pi_4(\tau^{(i-1)} \mid (\sigma^2)^{(i)}, \mu^{(i)}, \boldsymbol{Y}(\boldsymbol{t}), \boldsymbol{y})}\times \frac{\tau^\ast}{\tau^{(i-1)}}\right]
\end{equation}
and set $\tau^{(i)}$ to $\tau^{(i-1)}$ otherwise.  The proposal scale $\phi$ is adaptively set to produce an acceptance rate around 0.35. 


\subsection{Sensitivity analyses according to $k$, $m$, and the data generation assumption}\label{app3sec2}

First, we conduct a sensitivity analysis according to the various values of $k$ and $m$ of the Beta$(km, k(1-m))$ prior distribution on $\theta$. The setting is the same as that in Section~\ref{app2sec2}; in addition to the Uniform(0, 1) prior on $\theta$, we try $k=n$ and $k=n/5$, and three values of $m$, i.e., 0.01, 0.05, and 0.1. We fit the models on the simulated data, $\boldsymbol{y}^{\textrm{sim}}$.

\begin{figure}[t!]
\begin{center}
\includegraphics[scale=0.425]{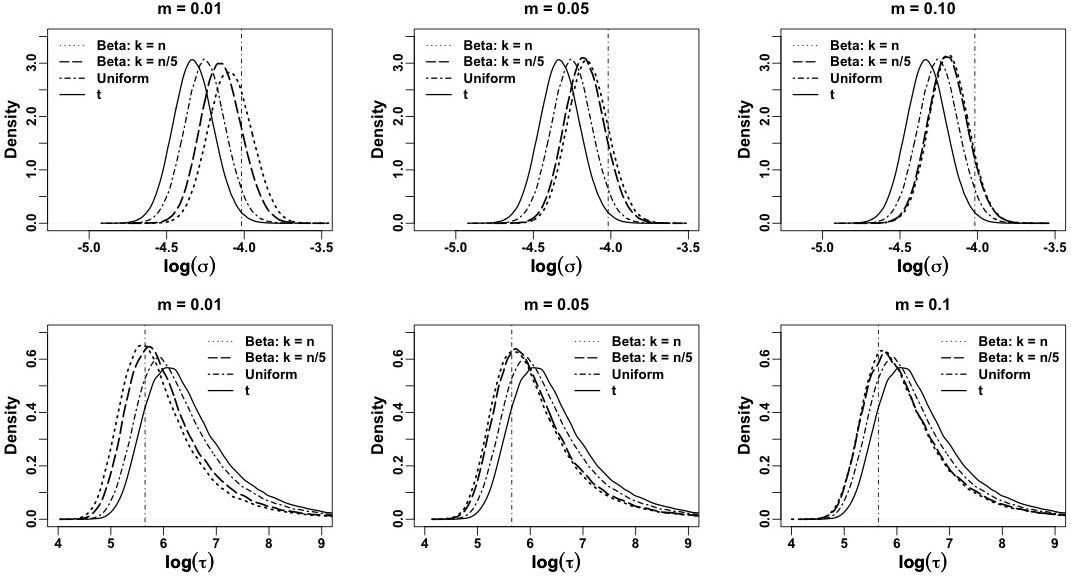}
\caption{The result of sensitivity analysis on $\log(\sigma)$ (first row) and $\log(\tau)$ (second row). Each panel shows four marginal posterior densities obtained by the $t_\nu$ error and proposed mixture error model with different priors on $\theta$. \textcolor{black}{The vertical dot-dashed lines indicate the generative true values.}  Clearly, the posterior densities obtained by the proposed mixture error model become similar to the density obtained by the $t_\nu$ error model as the Beta prior approaches the Uniform(0, 1) prior.}
\label{figure_app3}
\end{center}
\end{figure}

Figure~\ref{figure_app3} displays  the result. Each panel on the first row exhibits four marginal densities of $\log(\sigma)$ obtained by the $t_\nu$ error and proposed mixture error models (with different priors on $\theta$), and each panel on the second row shows those of $\log(\tau)$. Clearly, the marginal posterior density of the proposed mixture error model approaches the corresponding density of the $t_\nu$ error model as $k$ decreases or $m$ increases. It confirms again that as the Beta prior on $\theta$ becomes close to the Uniform(0, 1), the resulting inference of the proposed mixture error model becomes similar to that of the $t_\nu$ error model.

We also check the data generation assumption by  simulating  a new data set via $t_4$ errors instead of Gaussian errors. Given $\mu_{\textrm{gen}}=17.667$, $\sigma^2_{\textrm{gen}}=0.018^2$, and $\tau_{\textrm{gen}}=284.066$, we generate $\boldsymbol{Y}^{\textrm{sim}}(\boldsymbol{t})$ from (15) and  then generate $y_i^{\textrm{sim}}$ from a shifted and scale $t_4$ distribution, i.e, $Y^{\textrm{sim}}(t_i) + V^{0.5}_i t_4$ for all $i$. Using these newly simulated data, we repeat the sensitivity analysis, fitting the $t_\nu$ error and proposed mixture error models.

\begin{figure}[b!]
\begin{center}
\includegraphics[scale=0.425]{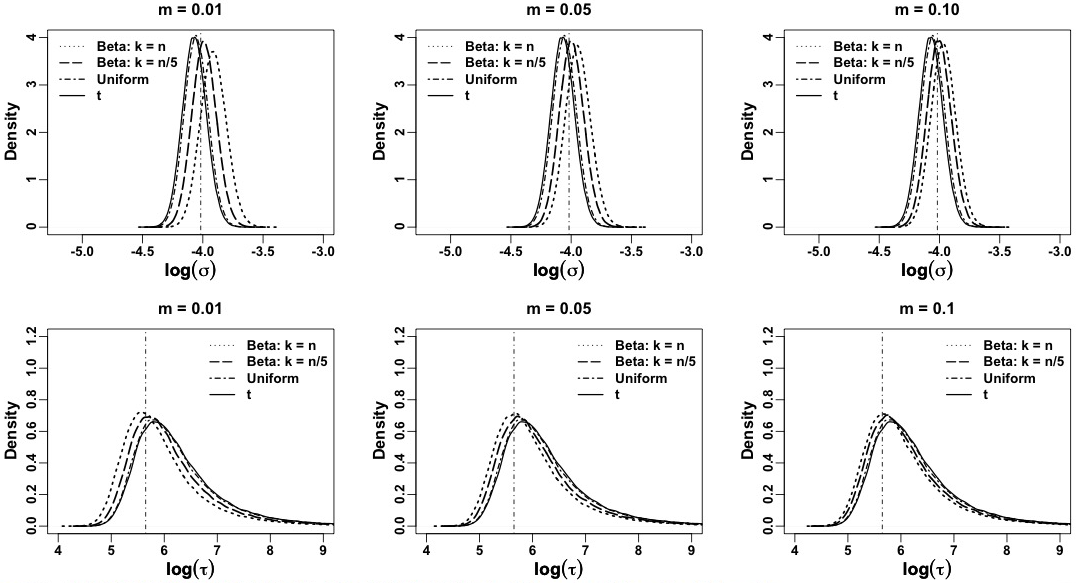}
\caption{The result of checking the sensitivity when a new data set is generated by $t_4$ errors instead of Gaussian errors. Each panel shows four marginal densities of $\log(\sigma)$ (first row) or $\log(\tau)$ (second row). \textcolor{black}{The vertical dot-dashed lines indicate the generative true values.}  It turns out that the density of $\log(\sigma)$ obtained by the $t_\nu$ error model results in the most accurate inference, while that of $\log(\tau)$ does not due to the negative association between $\sigma$ and $\tau$. However, the posterior distributions from the $t_\nu$ and mixture error models differ little, considering that the data are generated by $t_4$ errors.}
\label{figure_app3_5}
\end{center}
\end{figure}

Figure~\ref{figure_app3_5} displays the result of the sensitivity analysis. Regardless of the values of $m$, the $t_\nu$ error model produces a posterior distribution of $\log(\sigma)$ that concentrates more on $\log(\sigma_{\textrm{gen}})$ than the others, while that of $\log(\tau)$ does not  put more mass near $\log(\tau_{\textrm{gen}})$ than the others due to the negative association between $\sigma$ and $\tau$ a posteriori. Overall, the inference of the proposed mixture error model is similar to that of the $t_\nu$ error model,  considering that the data are generated by  $t_4$ errors.

\subsection{Markov chain convergence diagnostics}\label{app3sec3}

\begin{figure}[b!]
\begin{center}
\includegraphics[scale=0.425]{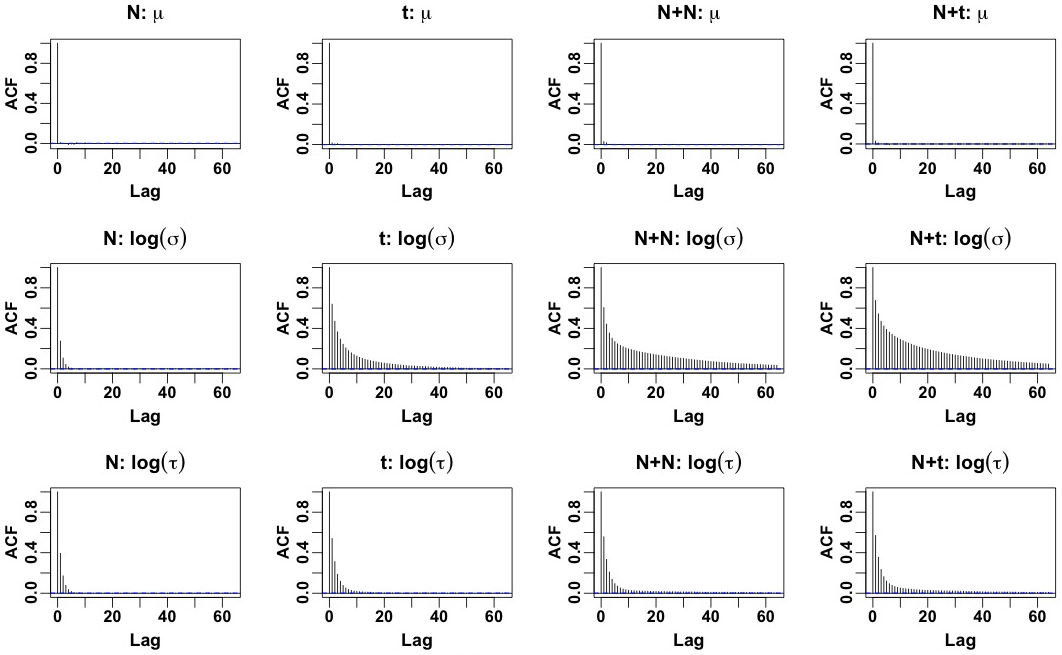}
\caption{Auto-correlation functions of $\mu$, $\log(\sigma)$, and $\log(\tau)$ obtained by fitting four different error models (N, $t_\nu$, N+N, N+$t_\nu$) on the simulated data $\boldsymbol{y}^{\textrm{sim}}$ used in Section~3.2.2. The auto-correlation functions decrease quickly, showing no evidence of the lack of convergence.}
\label{figure_app3_2}
\end{center}
\end{figure}

\begin{figure}[t!]
\begin{center}
\includegraphics[scale=0.425]{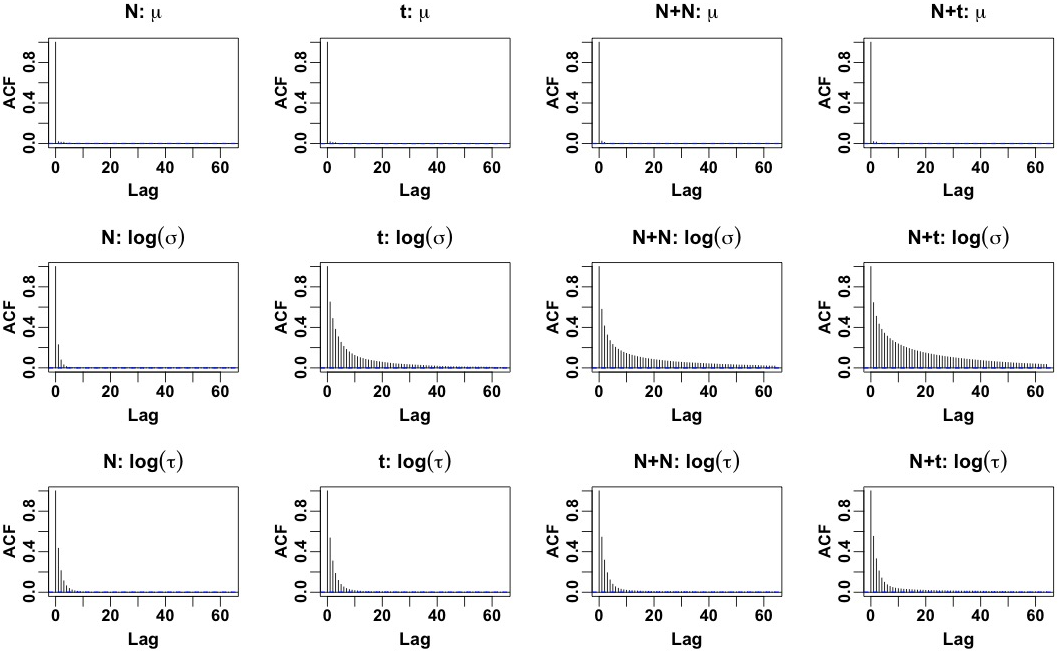}
\caption{Auto-correlation functions of $\mu$, $\log(\sigma)$, and $\log(\tau)$ obtained by fitting four different error models (N, $t_\nu$, N+N, N+$t_\nu$) on the MACHO data $\boldsymbol{y}$ used in Section~3.2.3. The convergence appears satisfactory, considering that all of the auto-correlation functions decrease quickly.}
\label{figure_app3_3}
\end{center}
\end{figure}

We check the convergence of the (thinned and combined) Markov chains used in Sections 3.2.2 (simulated data) and 3.2.3 (MACHO data). Figure~\ref{figure_app3_2} displays the auto-correlation functions of $\mu$ (first row), those of $\log(\sigma)$ (second row), and those of $\log(\tau)$ (third row) obtained by fitting four different error models on $\boldsymbol{y}^{\textrm{sim}}$, and Figure~\ref{figure_app3_3} shows those  fitted on the MACHO data $\boldsymbol{y}$. The auto-correlation functions decrease quickly for all cases. Also, though not shown here, the \textcolor{black}{ESSs of each combined posterior sample of $\mu$ for the Gaussian, $t_\nu$, Gaussian mixture, and proposed mixture error models} do not show the evidence of the lack of convergence.

\begin{figure}[t!]
\begin{center}
\includegraphics[scale=0.425]{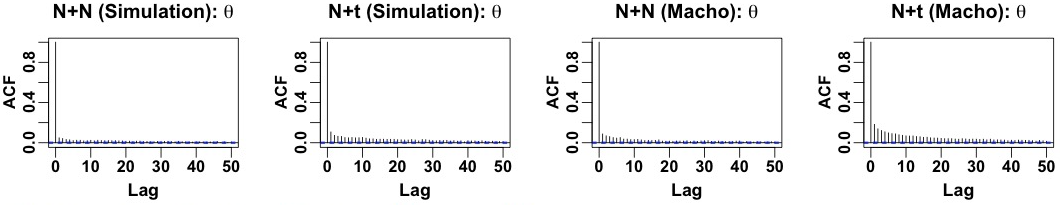}
\caption{Auto-correlation functions of $\theta$ obtained by fitting Gaussian (N+N) and proposed (N+$t_\nu$) error models on $\boldsymbol{y}^{\textrm{sim}}$ (the first two panels) and on  $\boldsymbol{y}$ (the last two panels). All of the auto-correlation functions decrease quickly, although the auto-correlation function of the proposed mixture error model decreases more slowly than that of the Gaussian mixture error model.}
\label{figure_app3_4}
\end{center}
\end{figure}

Figure~\ref{figure_app3_4} displays the auto-correlation functions of $\theta$ obtained by the Gaussian and proposed mixture error models fitted on both $\boldsymbol{y}^{\textrm{sim}}$ and $\boldsymbol{y}$. All of the auto-correlation functions decrease quickly. Their \textcolor{black}{ESSs per iteration} are 0.318, 0.217, 0.341, and 0.147 from the left, and \textcolor{black}{the ESSs per second} are 1977, 1204, 2245, and 814 \textcolor{black}{for the four models, respectively}.
\end{appendices}

\end{document}